\documentclass[twocolumn,prx,showpacs,notitlepage,nofootinbib,floatfix]{revtex4-1} 
\usepackage[utf8]{inputenc}

\usepackage{float}
\usepackage{amsmath,amssymb,setspace}
\usepackage{graphicx}
\usepackage{subfigure}  
\usepackage[usenames]{color}
\usepackage{graphicx}
\usepackage{bm, bbm}      
\usepackage{color}
\usepackage{srcltx}
\usepackage{upgreek}
\usepackage{slashed}

\usepackage{algorithmicx}
\usepackage[ruled,vlined]{algorithm2e}
\SetAlgoCaptionSeparator{ }

\newenvironment{algo}[1][!h]
  {
   \begin{algorithm}[#1]%
  }{\end{algorithm}}

\usepackage{xcolor}

\usepackage{bm, bbm}      
\usepackage{color}
\usepackage{srcltx}

\usepackage{slashed}

\def\=d{\, {\buildrel \rm def  \over =} \,}

\def\sqr#1#2{{\vcenter{\vbox{\hrule height.#2pt \hbox{\vrule width.#2pt height#1pt \kern#1pt \vrule width.#2pt}\hrule height.#2pt}}}}

\def\beq#1{\begin{equation} \label{#1}}
\def\eeq{\end{equation}}
\def\ben{\begin{equation*}}
\def\een{\end{equation*}}
\def\bequa{\begin{eqnarray}}
\def\eequa{\end{eqnarray}}

\def\bf#1{\bm{#1}}

\newcommand\bea{\begin{eqnarray}}
\newcommand\eea{\end{eqnarray}}

\def\beq{\begin{equation}}
\def\eeq{\end{equation}}

\def\al{\alpha}

\def\etol{\epsilon_{\text{tol}}}

\newcommand{\ad}{a^\dagger}

\newcommand{\van}{VanQver}
\newcommand{\hini}{H_{\text{ini}}}
\newcommand{\hinter}{H_{\text{nav}}}
\newcommand{\hfin}{H_{\text{fin}}}
\newcommand{\ang}{\overset{\circ}{\text{A}}}

\begin{document}

\title{VanQver: The Variational and Adiabatically Navigated Quantum Eigensolver}

\author{Shunji Matsuura}
\email{shunji.matsuura@1qbit.com}
\affiliation{1QB Information Technologies (1QBit), Vancouver, British Columbia, V6C 2B5, Canada}

\author{Takeshi Yamazaki}
\affiliation{1QB Information Technologies (1QBit), Vancouver, British Columbia, V6C 2B5, Canada}

\author{Valentin Senicourt}
\affiliation{1QB Information Technologies (1QBit), Vancouver, British Columbia, V6C 2B5, Canada}

\author{Lee Huntington}
\affiliation{1QB Information Technologies (1QBit), Vancouver, British Columbia, V6C 2B5, Canada}

\author{Arman Zaribafiyan} 
\affiliation{1QB Information Technologies (1QBit), Vancouver, British Columbia, V6C 2B5, Canada}


\begin{abstract}
    \begin{center}
        \vspace{-1.7em}
        (Dated: \today)
    \end{center}
The accelerated progress in manufacturing noisy intermediate-scale quantum (NISQ) computing hardware  has opened the possibility of exploring its application in transforming approaches to solving computationally challenging problems. The important limitations common among all NISQ computing technologies are the absence of error correction and the short coherence time, which limit the computational power of these systems.
Shortening the required time of a single run of a quantum algorithm is essential for reducing environment-induced errors and for the efficiency of the computation.
We have investigated the ability of a variational version of adiabatic quantum computation (AQC) to generate an accurate state 
more efficiently compared to existing adiabatic methods.
The standard AQC method uses a time-dependent Hamiltonian, connecting the initial Hamiltonian with the final Hamiltonian.
In the current approach, a navigator Hamiltonian is introduced which has a non-zero amplitude only in the middle of the annealing process.
Both the initial and navigator Hamiltonians are determined using variational methods.
A hermitian cluster operator, inspired by coupled-cluster theory and truncated to single and double excitations/de-excitations, is used as a navigator Hamiltonian.
A comparative study of our variational algorithm (\van) with that of standard AQC, starting with a Hartree--Fock Hamiltonian, is presented. The results indicate that the introduction of the navigator Hamiltonian significantly improves the annealing time required to achieve chemical accuracy by two to three orders of magnitude. The efficiency of the method is demonstrated in the ground-state energy estimation of molecular systems, namely,  H$_2$, P4, and LiH.
\end{abstract}

\maketitle

\section{Introduction}
Due to the inherent many-body nature of quantum systems, obtaining energetically stable quantum states is one of the most difficult problems in computational chemistry and physics.
Despite decades of advancements in classical hardware and algorithms for simulating quantum systems, many important problems, such as computing accurate electronic correlation energies
for strongly correlated systems
and predicting chemical reaction rates, remain largely unsolved. In order to obtain accurate results, highly precise numerical methods are required, such as full configuration interaction (FCI) and coupled-cluster theory. The computational resources required to run these precise methods grows with the system size to the extent that even state-of-art supercomputers can handle only small-sized problems~\cite{Head-Gordon-2008-58}. Researchers have attempted to alleviate this issue by introducing heuristics and approximation techniques like problem decomposition methods to reduce the computational complexity of this problem. This establishes a trade-off between the accuracy of the approximate solution and the computational efficiency. In addition, a lot of effort has been put towards the exploration of other paradigms of computation. Quantum computing, for example, is a promising approach to mitigating this problem~\cite{Feynman:1982,Lloyd1073}.
There has been a recent increase in the number of experiments simulating quantum systems on quantum devices~\cite{PeruzzoPhotonicPQ,PhysRevLett.104.030502,PhysRevA.95.020501,doi:10.1021/acsnano.5b01651,Santagatieaap9646,GoogleHydrogen,Hardware-efficientIBM,1712.05771,180310238,180302047}.
The difficulty faced in these experiments is in executing operations without losing relevant quantum coherence.
Whereas quantum error correction will make it possible to perform an unlimited number of operations, the required resources are large, well beyond the capabilities of current hardware.
 Therefore, the development of methods that require less-stringent quantum coherence is essential for near-term quantum devices. A subset of the authors of the present work have previously investigated the idea of combining problem decomposition techniques in conjunction with quantum computing approaches~\cite{Yamazaki:1806.01305}.
Quantum--classical hybrid algorithms, such as the variational quantum eigensolver \mbox{(VQE)~\cite{Peruzzo:2014aa, VQE2, VQE3}}, 
are suitable from this perspective.
In addition to the algorithm requiring a shorter coherence time, the VQE has demonstrated robustness against systematic control \mbox{errors~\cite{Santagatieaap9646,GoogleHydrogen,Hardware-efficientIBM}}.

Thus far, most of the experiments that have made use of the VQE and phase estimation algorithms \mbox{(PEA)~\cite{AbramsLloyd:PEA1997, Aspuru-Guzik:2005aa}} have been performed within the framework of gate model quantum computation.
An alternative framework is adiabatic quantum computation (AQC)~\cite{finnila_quantum_1994,kadowaki_quantum_1998,Farhi:01,Brooke1999,brooke_tunable_2001,Santoro,RevModPhys.80.1061,Farhi:01}. 
AQC solves computational problems by continuously evolving a Hamiltonian.
As such, it is absent of algorithmic errors (e.g, Trotterization errors). It is also robust against certain types of \mbox{decoherence~\cite{Albash:2015nx,childs_robustness_2001,PhysRevLett.95.250503,quant-ph/0507010,PhysRevA.71.032330}}.
Whereas gate model quantum computation is not meaningfully executed beyond the qubit dephasing time~\cite{Aharonov:96a}, the annealing time $T$ in AQC can be longer than the qubit coherence time under certain conditions.
For instance, when the interaction between the quantum system and environment is weak,
the decoherence occurs in the energy eigenbasis of the system.
In this case, the coherence of the instantaneous ground state, relevant for AQC, is preserved~\cite{Albash:2015nx}.
Nevertheless, the long annealing time in open quantum systems can cause problems.
For instance, it induces thermalization and the probability of finding a ground state decays exponentially at a fixed temperature
as the problem size increases~\cite{TScalingLaw}.
Therefore, effort towards  shortening the annealing time is essential and critical for the success of AQC.
The computational time in AQC is constrained by various factors.
From the perspective of computational efficiency and the prevention of bath-induced errors, 
a shorter annealing time is desirable, whereas, 
in order to avoid errors due to non-adiabatic transitions, the annealing time needs to be longer than
the scale of the inverse energy gap between the ground state and excited states during the annealing process.
In this work, we investigate a variational AQC method and demonstrate that it can significantly reduce annealing time.

\section{The Variational and Adiabatically Navigated Quantum Eigensolver}
%
In the standard form of AQC, the time-dependent Hamiltonian is given by
\bea
H(t)=A(t)\hini+B(t)\hfin
\label{standard time dep H}\,,
\eea
where the functions $A(t)$ and $B(t)$ satisfy the conditions that
\mbox{$A(0)\gg B(0)=0$ 
and  $B(T)\gg A(T)=0$}, respectively. 
Here, $T$ is the annealing time and $t\in[0,T]$.
Hereafter, we use the phrase ``standard AQC" for (\ref{standard time dep H}) with the boundary conditions for $A(t)$ and $B(t)$.

In the variational approach, we introduce a navigator Hamiltonian that has a non-zero amplitude only during the annealing process.
This navigator Hamiltonian, $\hinter(\bf{\theta} )$, is characterized by variational parameters $\bf{\theta}$.
Adding an extra term to the Hamiltonian has previously been considered (see Refs.~\cite{Farhi:differentPaths,Perdomo-Ortiz:2011,crosson2014different,VQE3}).
We take advantage of the fact that the initial Hamiltonian $\hini$ is not uniquely determined, and treat its parameters as variational: $\hini (\bf{\eta} )$.
The variational parameters $\bf{\eta}$ may need to satisfy certain constraints. We will address this aspect with concrete molecular models in Sec.~\ref{sec:Molecular systems}.
Note that even though different values of $\bf{\eta}$ may have the same initial ground state, they generate different quantum states during annealing.
A motivation for introducing $\hini (\bf{\eta} )$ and $\hinter(\bf{\theta} )$ with variational parameters is that
the efficiency of AQC can be highly dependent on the annealing paths.
The problem is that we do not \mbox{\textit{a priori}} know what kinds of terms would be beneficial for specific problems.
The variational parameters in AQC navigate the quantum state along the path that provides a higher probability 
of finding the true ground state, even when the annealing time is shorter than expected from standard AQC~(\ref{standard time dep H}).
We call this algorithm the ``Variational and Adiabatically Navigated Quantum Eigensolver''  (\van).
The time-dependent Hamiltonian in \van~is given by
\bea
H_{(\bf{\eta} ,\bf{\theta})}(t)=A(t)\hini (\bf{\eta} )+B(t)\hfin+C(t)\hinter(\bf{\theta} )\,.~~~
\label{vanQ Hamiltonian}
\eea
The coefficient $C(t)$ satisfies the boundary conditions
$C(0)=C(T)=0$, while its value is non-zero  during annealing, $0<t<T$.
A natural choice of the boundary conditions for $A$ and $B$ will be the same as that in standard AQC.
There can be multiple initial and navigator Hamiltonians, each with a different time-dependence.
At the end of the annealing process, the quantum device generates a certain state. 
We measure the expectation value of the final Hamiltonian $\hfin$.
If $\hfin$ is a quantum Hamiltonian, single-qubit rotations may be needed prior to performing measurements.
We send the data of the expectation value $E=\langle \hfin \rangle$ and the variational parameters $\bf{\theta}$ and $\bf{\eta} $
to a classical computer, where a classical optimizer will return a new set of variational parameters.
With the new set of parameters, we run AQC on the quantum device and then measure the energy.
We repeat this cycle until the energy converges.
The process is summarized in Fig.~\ref{fig:VanQver}. More-detailed steps of this algorithm are provided in  Appendix~\ref{App:algorithm}.
\begin{figure*}[ht]
\subfigure[]{\includegraphics[scale=0.55]{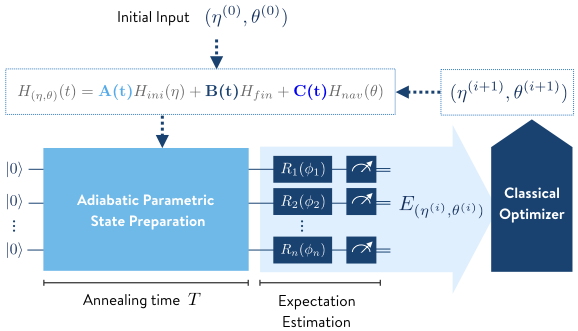}  \label{fig:VanQver4}}
\subfigure[]{\includegraphics[scale=0.3]{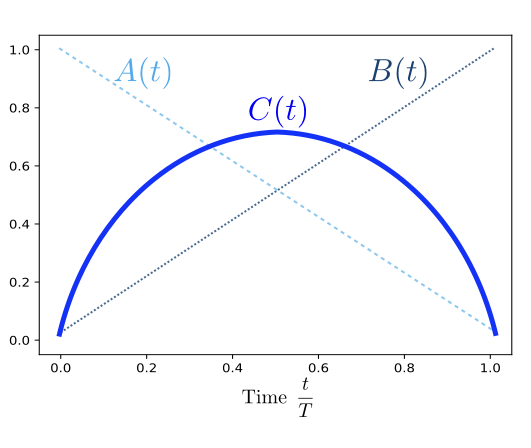} \label{fig:VanQver5}}
\caption{(a) A schematic illustration of the quantum--classical hybrid  \van~algorithm, and (b) a schematic illustration of the time dependent amplitudes of $\hini,\hinter$, and $\hfin$.  }
 \label{fig:VanQver}
\end{figure*}

In this work, we focus on investigating \van~in the context of quantum chemistry simulations.
However, the use of the algorithm itself can be much more general.
For instance, in solving combinatorial optimization problems,
one could view \van~as a refinement and a unification of previously studied techniques, such as the use of an
anti-ferromagnetic driver Hamiltonian (i.e., a nonstoquastic Hamiltonian)~\cite{farhi:12,
Seoane:2012uq,crosson2014different,Seki:2015,Zeng:2016bs,Hormozi:2016aa}, the use of an
inhomogeneous driver Hamiltonian~\cite{Inhomo-DicksonAmin,InhomoDriver,Inhomo-pspin}, 
and reverse annealing~\cite{Perdomo-Ortiz:2011,ReverseAnnealingONL}.
Reverse annealing is a method for  annealing backwards from a particular state by increasing quantum fluctuations and then reducing them in order to reach a new state.
It has been implemented on D-Wave Systems' quantum annealer~\cite{ReverseAnnealingONL,DwaveReversAnnealTopological}.
In~\cite{Perdomo-Ortiz:2011}, a uniform transverse field is considered as an additional Hamiltonian with a ``sombrero-like'' time-dependent amplitude (similar to $\hinter$ in \van, but with different functionality),
while the initial Hamiltonian is diagonal in the computational basis and updated iteratively with heuristic guesses. Whereas these specific methods can improve computational results in certain cases, 
guidelines for applying them to general models are needed. 
In contrast with the works cited above, \mbox{\van} keeps a variational transverse field as the driver Hamiltonian while adding a prominence-like navigator Hamiltonian. This algorithm constitutes a blueprint for one approach to solving this problem.

\section{Molecular Systems}
\label{sec:Molecular systems}
Let us consider the problem of obtaining the ground state energy of a molecule. 
Within the Born--Oppenheimer approximation, the second quantized form of the molecular electronic Hamiltonian is obtained as
\bea
\hfin^{\text{ferm}}=\sum_{pq}h_{pq}a^{\dagger}_{p}a_{q}+{1\over 2}\sum_{pqrs}h_{pqrs}a^{\dagger}_{p}a^{\dagger}_{q}a_{r}a_{s}\,,
\label{fermionic Hamil}
\eea
where $p,q,r$, and $s$ label spin-orbitals and 
$\ad_p$ and $a_p$ are creation and annihilation operators of an electron in spin-orbital $p$, while
$h_{pq}$ and $h_{pqrs}$ are one- and two-electron integrals. 
Under Bravyi--Kitaev (BK) \cite{Bravyi:00} or Jordan--Wigner (JW) \cite{Jordan1928} transformations, Eq.~(\ref{fermionic Hamil}) is translated into a
qubit Hamiltonian $\hfin^{\text{qubit}}$.

In general, the obtained  $\hfin^{\text{qubit}}$ has the form of a $k$-local Hamiltonian with $k\ge3$.
$k$ scales as $\mathcal{O}(N)$ for the JW transformation whereas the scaling is $\mathcal{O}(log N)$ for the BK transformation. Here, $N$ is the number of qubits.
While $k$-local qubit Hamiltonians are directly implemented on a classical hardware in our numerical calculations, it is important to 
mention that $k$ is limited to $\mathcal{O}(1)$ on actual quantum devices.
A standard method to circumvent this problem is to use perturbative gadgets
\cite{kempe:1070,Jordan:08}.
In this case, BK transformation requires less overhead in computational resources. See, for instance,
\cite{AQC-for-qchem,Bravyi-KitaevSeeley,Tranter-BravyiKitaev} for quantum chemistry simulation in BK transformation.

A natural initial Hamiltonian consists of one-electron terms,
which include the \mbox{Hartree--Fock (HF)} Hamiltonian.
For simplicity, we use a 1-local qubit Hamiltonian in the rest of our \mbox{experiments}: 
\bea 
\hini^{\text{qubit}}=\sum_{p}{\eta_{p}}\sigma^{z}_{p}\,.
\eea
While the ${\eta_{p}}$ takes specific values in the HF Hamiltonian, 
 we treat them as variational parameters in \mbox{\van}. 
However, there is an important symmetry constraint.
The signs of ${\eta_{p}}$ determine which spin-orbitals are occupied or virtual.
Since the electron number operator ${N}$ and spin number operators ${N}_{\uparrow, \downarrow}$, or a parity of them, commute with 
 $\hini$, $\hfin$, and the cluster operators ($\hinter$), these numbers are constant during annealing.
 Therefore, the signs of ${\eta_{p}}$ directly determine the state of the molecule, such as being neutral or ionic, at the end of the annealing process.
 We vary the parameters ${\eta_{p}}$ while keeping their signs fixed.

The choice of the navigator Hamiltonian is important for the efficiency of \van.
Inspired by the promising results of VQE using the unitary coupled-cluster (UCC) ansatz, we propose to use a hermitian
cluster operator as the navigator Hamiltonian in our molecular simulations.

The cluster operator truncated to single and double excitations is given by
\bea
\hspace{-1em}
\hinter^{\text{ferm}}=\sum_{i\in \text{occ}\atop \al \in \text{vir}}\theta_{i\al}a^{\dagger}_{i}a_{\al}+\sum_{ij\in \text{occ}\atop\al\beta \in \text{vir}}\theta_{ij\al\beta}a^{\dagger}_{i}a^{\dagger}_{j}a_{\al}a_{\beta}+h.c.\,,
\label{uccsd Hamil}
\eea
where $h.c.$ is the Hermitian conjugate.
One can add higher-order excitation terms as well. In VQE, the entire quantum operation is a realization of the Trotterized exponential of an anti-Hermitian cluster operator, and the accuracy of the obtained energy is limited by which excitations are included. 
For instance, an exact state can be obtained if all possible excitation operators are included, whereas 
the use of only single excitation operators with the HF initial state may not significantly improve the performance 
due to
Brillouin's theorem, which states that the HF ground state cannot be improved by mixing it with singly excited determinants.
In \van, a quantum state can, in principle, reach the exact state without the cluster operator.
Therefore, the accuracy is not limited by the type of excitation operators in $\hinter$.
The role of a cluster operator is to {\it assist} in reaching the exact state as closely as possible within a shorter annealing time $T$.

\section{Numerical Results}
The performance of \van{ }was tested by solving the time-dependent Schr{\"o}dinger equation
directly on classical hardware. 
The calculations were performed using the library QuTip~\cite{qutip1,qutip2}.
The annealing schedule used was
$A(t)=1-\left({t\over T}\right)^2, B(t)=\left({t\over T}\right)^2$ and $C(t)=\al {t\over T}\left(1-{t\over T}\right)$,
where $\al$ is a numerical constant which
 can be renormalized to 1 by rescaling $\bf{\theta}$.
The initial parameters $\bf{\theta}^{(0)}$ were set to  zero and $\bf{\eta}^{(0)}=\text{sign}({\bf \eta}_{\text{HF}})$,
where ${\bf \eta}_{\text{HF}}$ are the coefficients in the HF Hamiltonian.
All the parameters were updated using an optimizer on a classical computer. 
The test set included H$_2$, P4 (two hydrogen molecules parallel to each other) for various values of the separation distance $d$, and LiH.
The graphical picture of the molecules is given in Appendix \ref{App:geometries of molecules}.
The nuclear separation distances for both H$_2$ and LiH were chosen to be \mbox{$1$ $\ang$}.
For P4, the nuclear separation distance for each of the hydrogen molecules was \mbox{$2$ $\ang$}, and the separation distance $d$ between the two hydrogen molecules
varied from \mbox{$d=0.4$ $\ang$} to $d=4.0$ $\ang$. 
In what follows, units of distance are always expressed in  ${\mathring{\textrm{A}}}$ngstr{\"o}ms. Also note that a minimal basis set (STO-3G) is employed in all calculations. In this case, H$_2$, P4, and LiH were described using 4, 8, and 12 qubits, respectively.
Note that the number of qubits can be reduced based on 
 $\mathbb{Z}_2$ symmetries~\cite{TaperingOffQubits,GoogleHydrogen}.

%
%
The Broyden--Fletcher--Goldfarb--Shanno (BFGS) algorithm was used as a classical optimizer and the tolerance for termination ($\epsilon_{\text{tol}}$) was set to \mbox{$\epsilon_{\text{tol}}=0.001$}, $0.0005$, and $0.0001$.
The converged parameters $(\bf{\eta}^{\text{final}},\bf{\theta}^{\text{final}})$
provided the annealing path, which brought the quantum state closest to the exact ground state in a given annealing time $T$.

Note that $T$ is the annealing time of a {\it single} run. The total computational time needs to take into account the repeated runs of the quantum device as well as the computational overhead for the classical optimizer. However, the limitation of near-term quantum hardware is the coherence time of a single run.
Therefore, the main focus was on the relation between the annealing time $T$ and the expectation value of the energy $E$ obtained using $\bf{\eta}^{\text{final}}$ and $\bf{\theta}^{\text{final}}$.

We compared the performance of \van~with the standard AQC (\ref{standard time dep H}).
The initial Hamiltonian was chosen to be a 
canonical RHF Hamiltonian, as considered in \mbox{Ref.~\cite{AQCHmp}},
\bea
H_{\text{MP}}&=&\sum_{p}f_{pp}a^{\dagger}_{p}a_{p},\cr
f_{pp}&=&h_{pp}+\sum_{i\in\text{occ}}(\langle pi | pi \rangle - \langle pi | ip \rangle  )\,,
\label{eq:Hmp}
\eea
where $h_{pp}$ are the one-electron integrals
and $\langle pi | pi \rangle$ and $ \langle pi | ip \rangle$ are direct (Coulomb) and exchange two-electron integrals, respectively.
Here, MP refers to M{\o}ller--Plesset, since this is the unperturbed (zeroth order) form of the Hamiltonian used in MP perturbation theory.
These terms of the Hamiltonian sum up to become the Fock operator.

\begin{figure}[ht]
\includegraphics[scale=0.48]{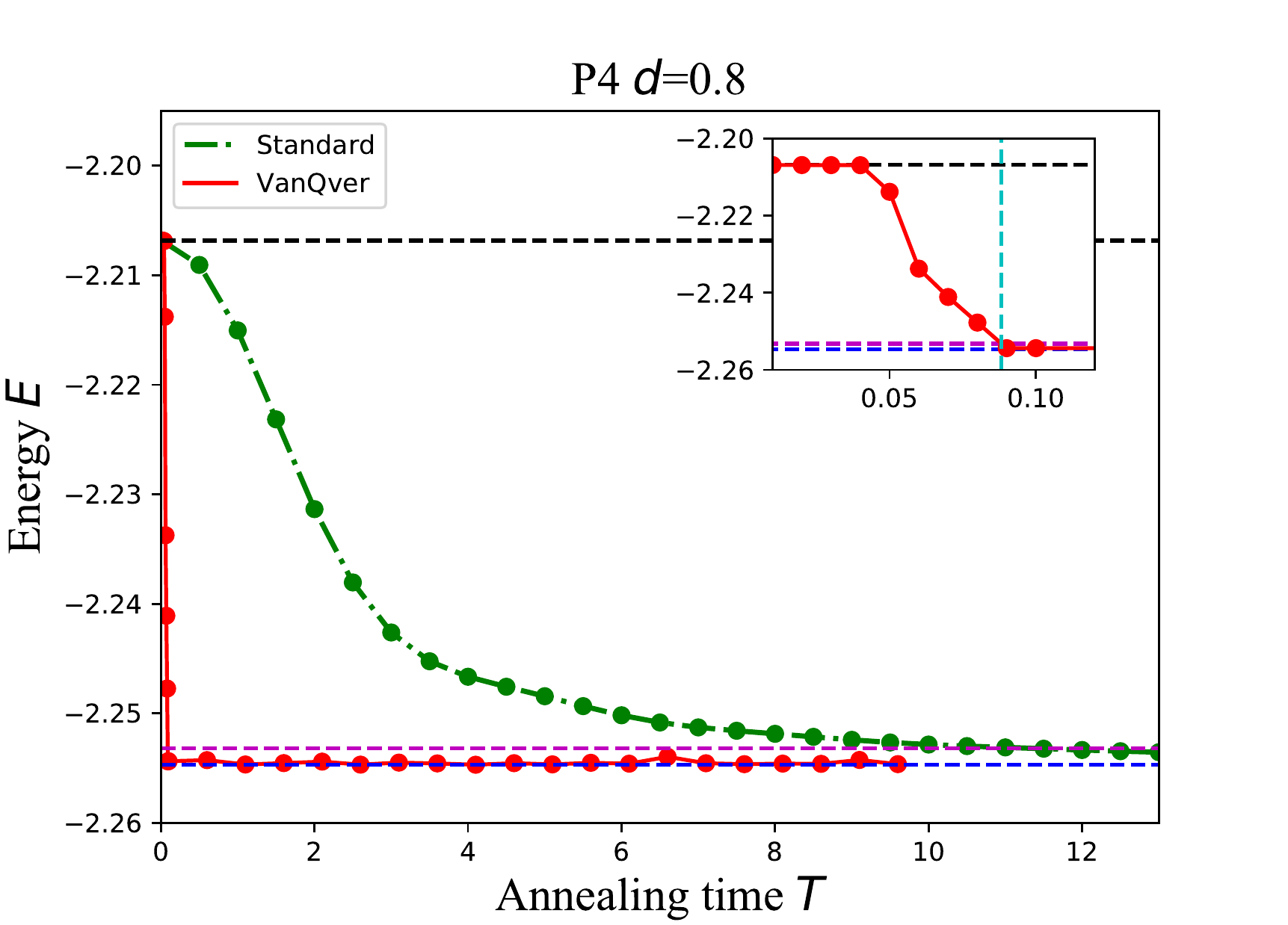} 
\caption{Energy of P4 with $d=0.8$ as a function of annealing time $T$.
The inset shows the region $0\le T\le 0.12$. 
The red points represent the results obtained using \van, the green points represent the results obtained with standard AQC with $\hini=H_{\text{MP}}$, and the dotted lines in blue and magenta dotted lines represent the
exact and chemical accuracy energies, respectively.
$T^{\text{\van}}_{CA}=0.088$ for \van, whereas $T^{\text{Standard}}_{CA}=11.5$ without using \van.
}
\label{fig:EvsT_P4-08}
\end{figure}

Fig.~\ref{fig:EvsT_P4-08} shows the obtained energy $E$ with $(\bf{\eta}^{\text{final}},\bf{\theta}^{\text{final}})$ 
as a function of the annealing time $T$ for the P4 molecule with separation distance $d = 0.8$ and  termination tolerance $\epsilon_{\text{tol}} = 0.001$. The energy is expressed in hartrees and the unit of the annealing time depends on the realization of the Hamiltonian on quantum hardware.
The time evolution used in the numerical simulation is $\mathcal{T}\exp(-i\int_{0}^{T}Hdt)$, where $\mathcal{T}$ is the time ordering.
Results for 
H$_2$ and LiH are shown in  Appendix~\ref{app:Numerical results for  H2 and LiH}.
%
%
%
The red points represent the results obtained with \van, while the green lines represent the results obtained with standard adiabatic evolution (\ref{standard time dep H}) when $H_{\text{MP}}$ (\ref{eq:Hmp}) was used for $\hini$.
The horizontal dotted line in blue corresponds to the exact energies, the horizontal dotted line in magenta shows the error from the exact energies within chemical accuracy (0.0015 Hartree), and the horizontal dotted line in black represents the Hartree--Fock energy.
The inset shows the very short annealing time region.
This result demonstrates that \van~allows us to reach chemical accuracy within a much shorter time compared
to the standard AQC time evolution (\ref{standard time dep H}).
The annealing time to reach chemical accuracy, $T_{CA}$, is 
$T_{CA}^{\text{\van}}=0.088$ 
 for \van,
whereas 
$T_{CA}^{\text{Standard}}=11.5$ when {\van} is not used.
We emphasize that $T_{CA}^{\text{\van/Standard}}$ is defined as the annealing time for a single run.
Another interesting point is that when the annealing time was too short, 
the energy 
did not differ significantly from the HF value and
the quantum states did not evolve very much from the initial state.
Therefore, the final state was still close to the reference state.
Once the annealing time became longer, the effect of $\hinter$ was very pronounced and the energy dropped rapidly.
Note that while the changes in $\bf{\eta}$ within the appropriate parameter region, did not affect the initial state,
it did change the evolution during the annealing process. Therefore, we varied $\bf{\eta}$ depending on the values of $\bf{\theta}$.

\begin{figure}[ht]
\includegraphics[scale=0.5]{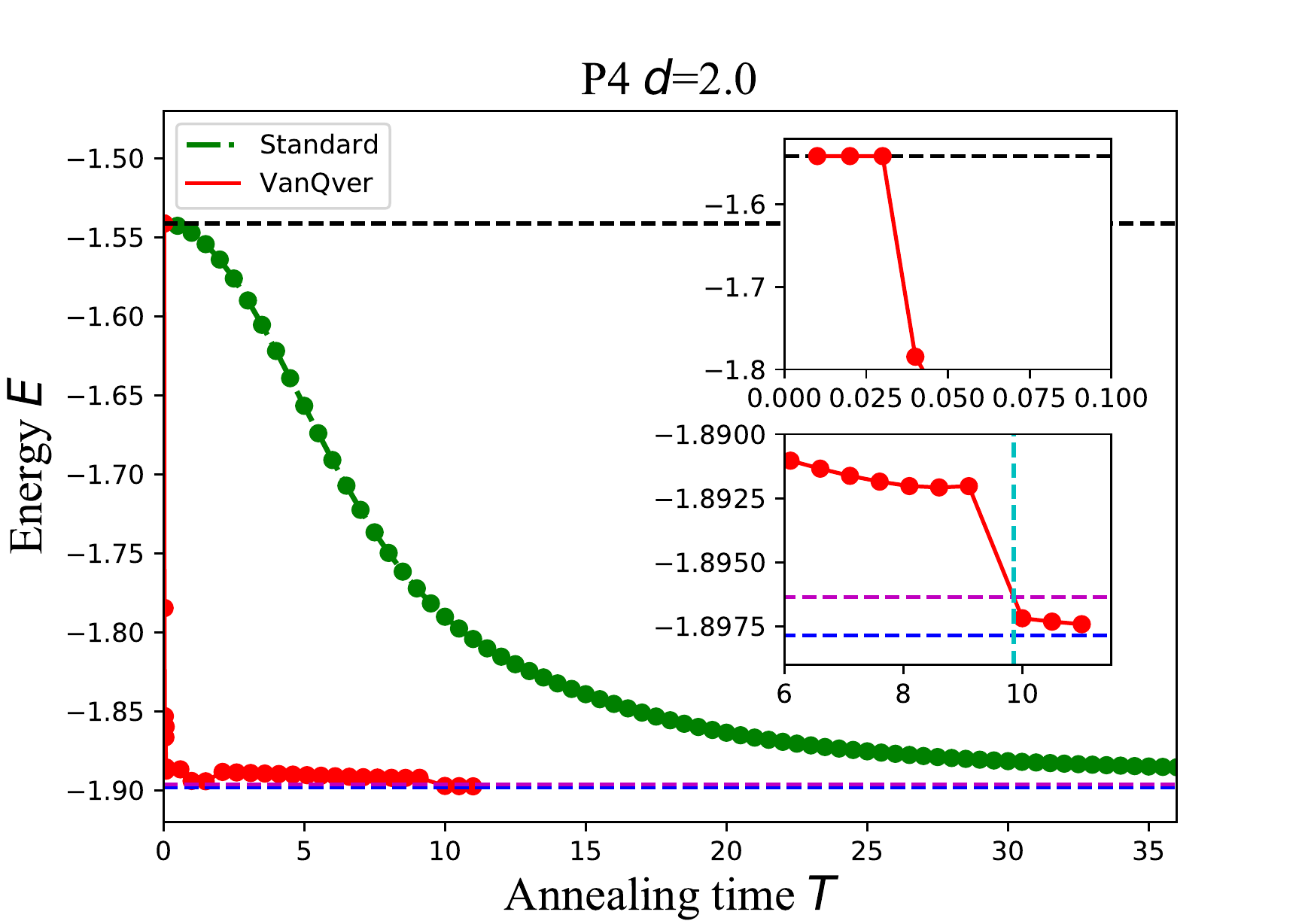}
\caption{
Energy as a function of annealing time $T$ for P4 with $d=2.0$. 
The exact energy (FCI energy) is $-1.8978$.
The energy drops rapidly from $T=0.03$ to $T=0.1$. The annealing times to reach chemical accuracy are
 $T^{\text{\van}}_{CA}=9.855327037$ for {\van} and
 $T^{\text{Standard}}_{CA}=456$ for standard AQC. The plot shows only $T\le36$.
The upper inset shows the time window in which the energy drops rapidly from the HF energy.
The lower inset shows the time window in which the energy achieves chemical accuracy.
}
\label{fig:EvsT_P4-20}
\end{figure}

Fig.~\ref{fig:EvsT_P4-20} shows the results for P4 with $d=2.0$. In this case, we observe slightly different features.
Similar to the case of P4 with $d=0.8$, the energy 
did not vary significantly from
the HF energy when the annealing time was too short, and then dropped rapidly with larger annealing times. In the case of $d=2.0$, the energy did not reach chemical accuracy immediately.
Instead, it decreased gradually after the rapid drop and then eventually reached chemical accuracy
at $T^{\text{\van}}_{CA}=9.855$.
The annealing time required to reach chemical accuracy was much longer than when $d=0.8$.
However, the required time using standard AQC was much longer: $T^{\text{Standard}}_{CA}=456$.
Therefore, once again, $T^{\text{\van}}_{CA}\ll T^{\text{Standard}}_{CA}$.

\begin{figure}[ht]
\includegraphics[scale=0.52]{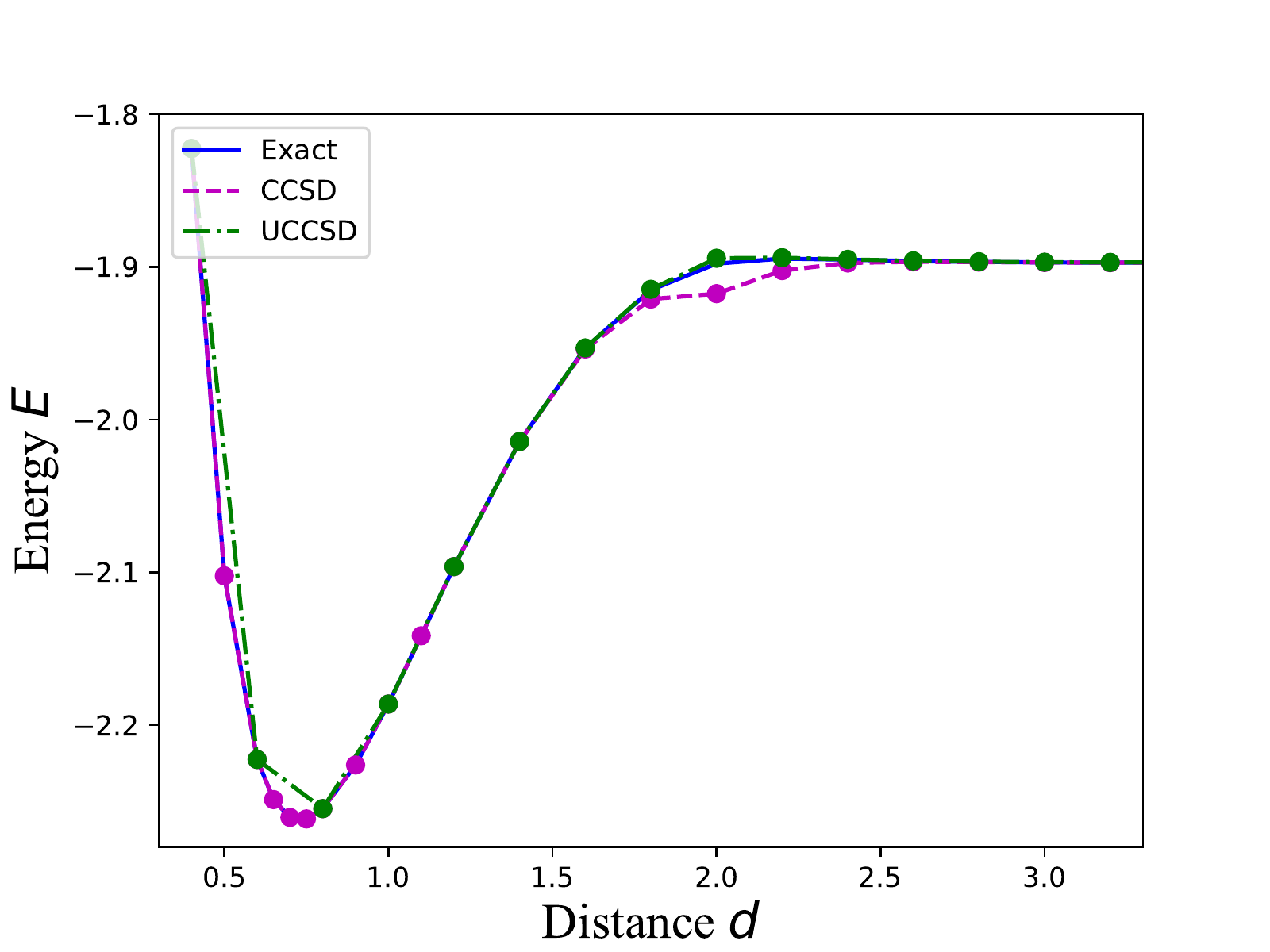}
\caption{Energy as a function of intermolecular separation distance for P4. The blue line represents the exact result, and the dotted line in magenta
and the dash-dotted line in green
represent the CCSD and the UCCSD results, respectively.
The Trotter number for UCCSD is 1.
}
\label{fig:EvsDistanceCCSD}
\end{figure}

\begin{figure}[ht]
\includegraphics[scale=0.45]{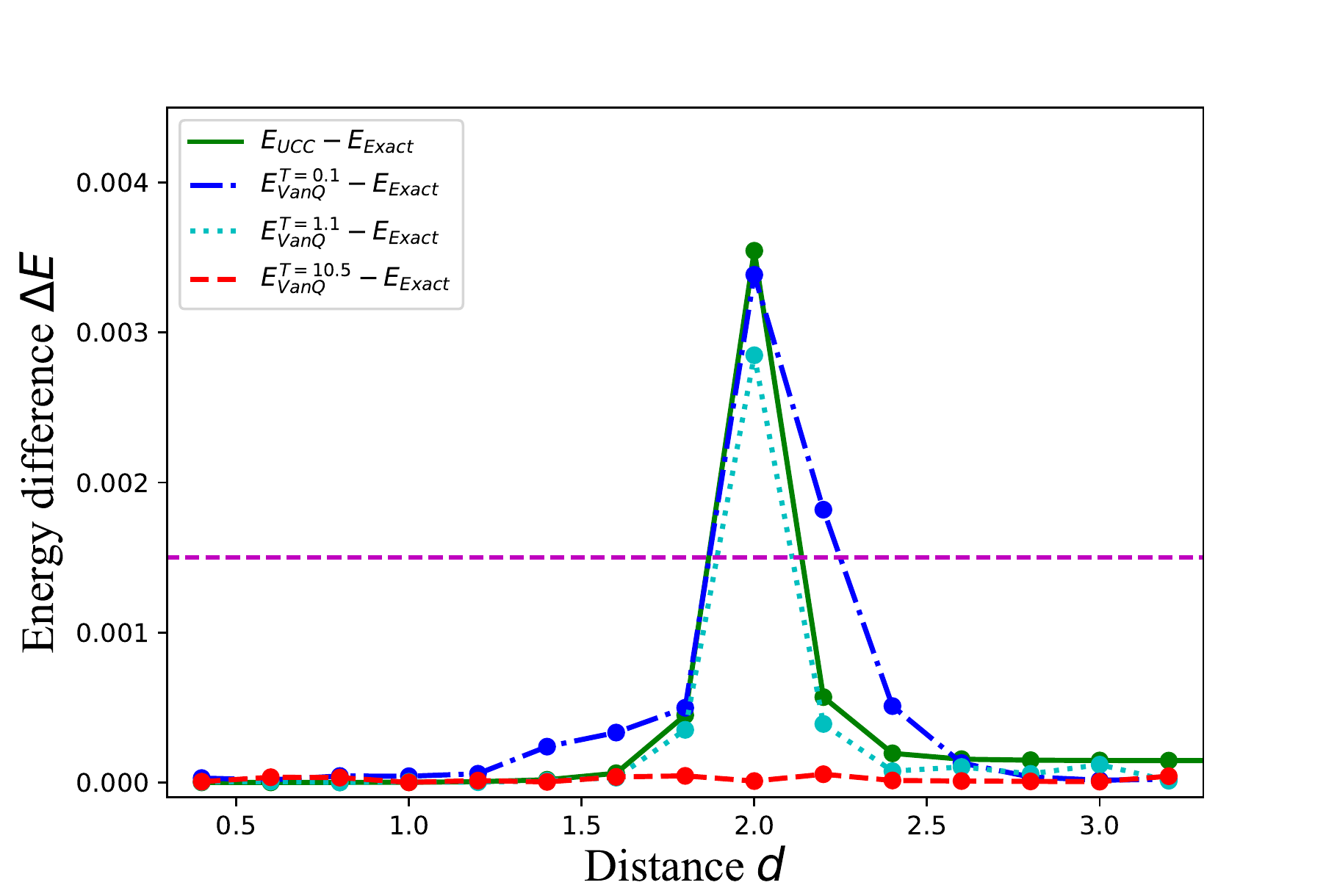}
\caption{The deviation of the energy in UCCSD and \van~from the exact results.
The annealing time is chosen to $T=0.1$ (the blue dash-dotted line),
$T=1.1$ (the cyan dotted line), and $T=10.5$ (the red dashed line).
The tolerance is $\epsilon_{\text{to;}}=0.0001$.
}
\label{fig:deltaEvsT_P4_UCC}
\end{figure}

We investigated the accuracy of \van~compared to conventional CCSD, as well as unitary CCSD (UCCSD) calculations on this system. 
The energy as a function of the separation distance for CCSD, UCCSD, and the exact results are shown in Fig.~\ref{fig:EvsDistanceCCSD}.
One can see that CCSD and UCCSD provide accurate results except near $d=2.0$.
Note that the latter is the accuracy achieved by the UCCSD circuit in VQE, in the absence of noise.
Since the scale of the chemical accuracy is much smaller than that of the the total energy, the energy differences between the UCCSD results and the exact results,
as well as between \van~results and the exact results are plotted in
Fig.\ref{fig:deltaEvsT_P4_UCC}.
This figure illustrates that the UCCSD ansatz in VQE fails to achieve chemical accuracy at $d=2.0$. 
In the case of \van, the accuracy at $d=2.0$ is about the same 
as that of UCCSD when the annealing time is very small ($T=0.1$
and $1.1$). As the annealing time becomes longer ($T=10.5$)
the energy difference becomes much smaller than 
the chemical accuracy (0.0015 Hartree).
This reflects the fact that the AQC is solving Full CI
and the accuracy is not restricted by the ansatz.
We emphasize that the residual energy for $T=10.5$
in the standard AQC is much larger than the chemical accuracy;
$\Delta E = E^{T=10.5}_{\text{Standard}}-E_{\text{Exact}}=0.100272
\gg 0.0015$.

As is well-known, in order to obtain accurate results for the P4 system both at and around $d = 2.0$ (i.e., a square geometry), we would need to add the quadruple excitations (exact) or use a multi-reference correlation approach. This is due to the fact that two configurations become degenerate as $d$ tends towards a value of 2.0, in this system. This is a well-known pathological multi-reference case demonstrating the failures of conventional single-reference methods like CCSD and UCCSD. This is also reflected in the longer annealing times of the AQC simulations, both with the standard method (using a canonical RHF Hamiltonian) and \van.

The accuracy of the energy for a given annealing time $T$ depends on the tolerance of the termination.
Even with short annealing times, it was possible to improve the value of the obtained energy as the tolerance $\etol$ changed from
$0.001$ to $0.0001$, in exchange for an increase in iterations. 
The tolerance dependence of the energy and the number of iterations are shown in Fig.~\ref{fig:EvsTol_P4-20} and 
Fig.~\ref{fig:ItevsTol_P4-20} 
in Appendix~\ref{Classical optimizer dependence}.


\begin{figure}[ht]
\includegraphics[scale=0.52]{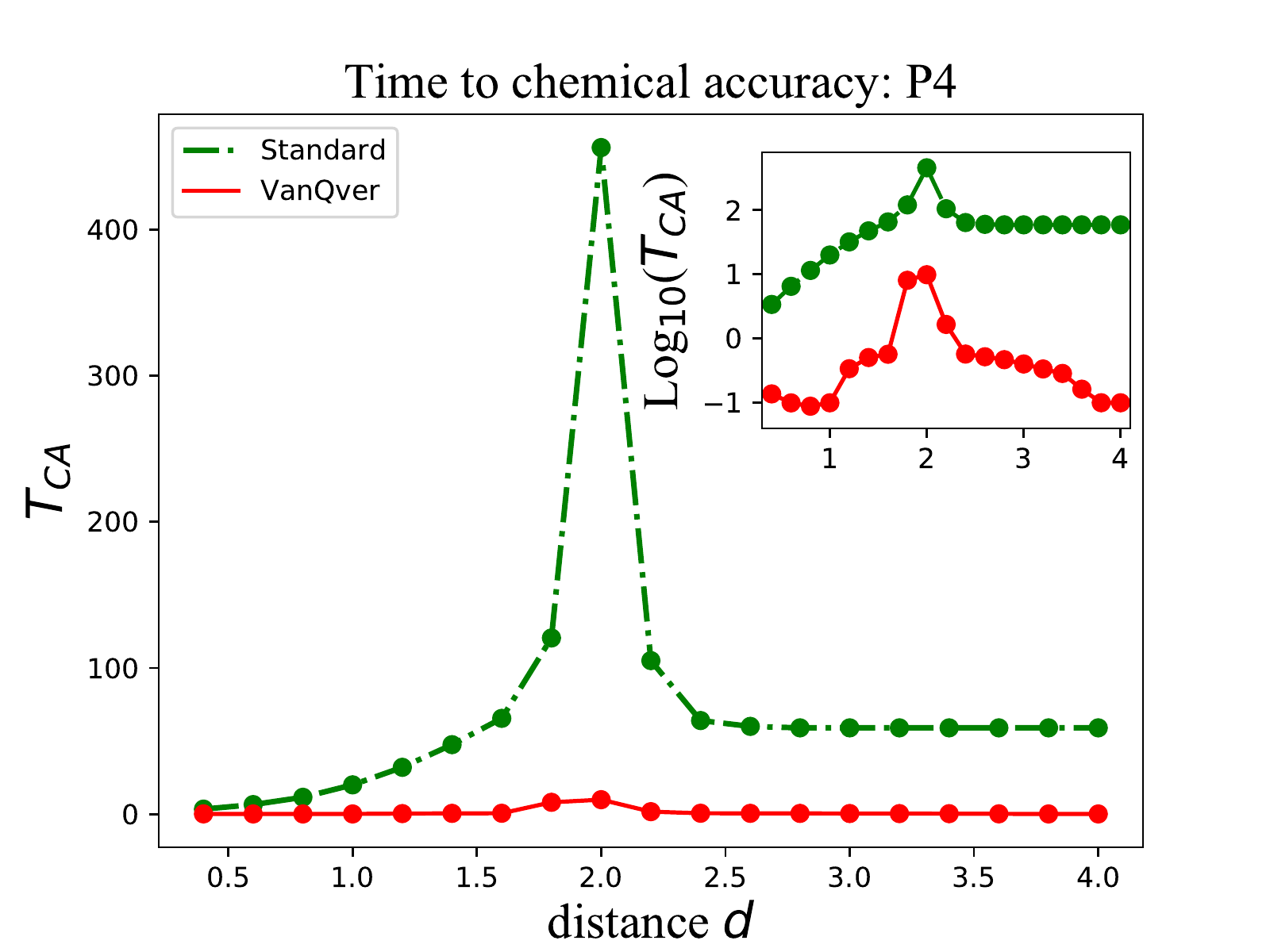}
\caption{The time to chemical accuracy $T_{\text{CA}}$ for P4. The horizontal axis corresponds to the intermolecular separation distance $d$. 
The red points represent the \van~results and the green points represent the standard AQC results 
with initial Hamiltonian $H_{\text{MP}}$. The time is represented on a logarithmic scale, in the inset. It shows that $T^{\text{\van}}_{\text{CA}}$ and
$T^{\text{Standard}}_{\text{CA}}$ differ by two to three orders of magnitude across the entire range of the intermolecular distance $d$. 
The tolerance is $\epsilon_{\text{to;}}=0.001$.
 }
\label{fig:TimeToCA}
\end{figure}
Fig.~\ref{fig:TimeToCA} shows a comparison of the annealing time to chemical accuracy for \van~and standard AQC for different distances of $d$ in P4. 
A logarithmic scale is used for $T_{CA}$ in the inset.
The general features of the two results are similar. When the required annealing time is longer in standard AQC, similar behaviour is observed in \van.
It is not surprising that the longest required annealing time is observed at an intermolecular separation of $d = 2.0$, since this is where the conventional CCSD method fails (Fig.~\ref{fig:EvsDistanceCCSD}), as discussed above. However, we would like to emphasize that the required annealing time is always one or two orders of magnitude shorter than that of standard AQC.

\section{Role of the Navigator Hamiltonian}
\label{Sec:Role of the Navigator Hamiltonian}

Computational results of AQC depend on various factors.
The original proposal of AQC uses adiabatic paths to reach the accurate ground state of $\hfin$.
For this purpose, the following adiabaticity condition needs to be satisfied:
\bea
T\gg {\Big|\langle \phi_{1}(s)|{dH(s)\over ds}|  \phi_{0}(s)\rangle\Big| \over
\Delta E(s)^2
} ,
\label{adiabatic cond}
\eea
where $s\equiv t/T$, $|\phi_{0}(t)\rangle$ and $|\phi_{1}(t)\rangle$ are the instantaneous ground and first excited states.
$\Delta E(t)$ is the energy gap between them.
This can be achieved by either choosing the annealing time $T$ to be large so that $H(t)$ changes sufficiently slowly, or finding an annealing path on which the energy gap stays open. 
In some cases, however, it is more efficient to 
use excited states in the middle of annealing via non-adiabatic transitions.
In \van, the algorithm itself determines the optimal way to find the accurate solution.

In this section, the role of $\hinter$ is investigated for molecular systems.
In particular, we look at the following two quantities: the energy gap between the instantaneous ground state and the first excited state,
and the overlap between the instantaneous ground state and the 
quantum state generated in the annealing.
The former is used in the adiabatic condition (\ref{adiabatic cond}).
The latter captures dynamical aspects of the annealing.

\begin{figure}[ht]
\includegraphics[scale=0.55]{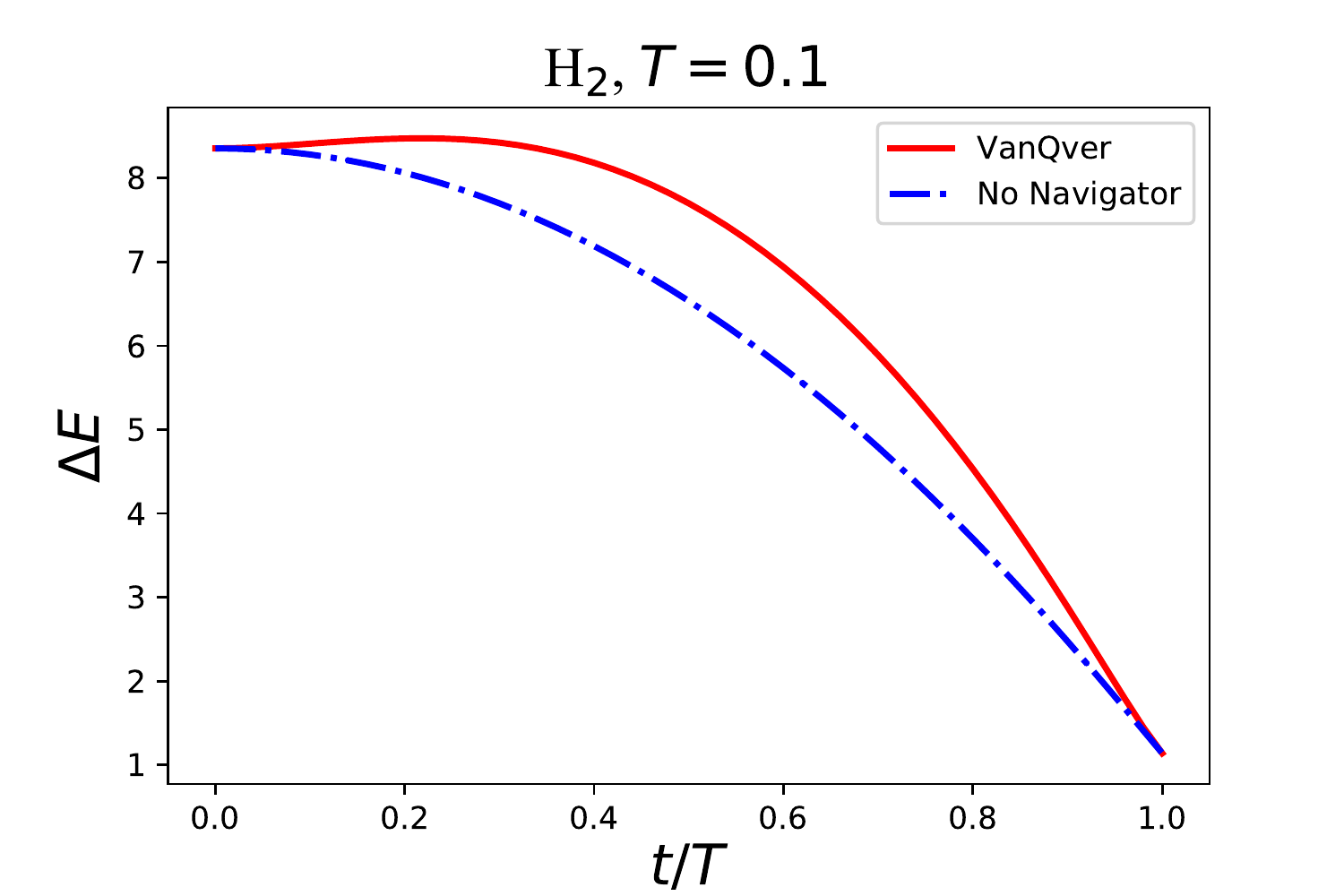}
\caption{Energy gap of the instantaneous ground state and the first excited state for the hydrogen molecule. The annealing time is $T=0.1$.
The red solid line represents the result of \van, and the blue dash-dotted line represents that without $\hinter$.}
\label{fig:gapvstH2}
\end{figure}

Let us take the optimal value of the variational parameters
$(\bf{\eta}^{\text{final}},\bf{\theta}^{\text{final}})$
for a given annealing time $T$. 
The $i$th excited instantaneous eigenstates $|\phi_i(t)\rangle$ and the eigenvalues $E_i(t)$ are defined as
\bea
H_{(\bf{\eta}^{\text{final}},\bf{\theta}^{\text{final}})}(t)|\phi_i(t)\rangle
=E_i(t)
|\phi_i(t)\rangle.
\eea
In order to understand the role of $\hinter$, we also consider a Hamiltonian 
without the navigator Hamiltonian $\hinter$, $\bf{\theta}=0$;
\bea
H_{(\bf{\eta}^{\text{final}},\bf{\theta}=\bf{0})}(t)|\phi^{\text{NoNav}}_i(t)\rangle
=E^{\text{NoNav}}_i(t)
|\phi^{\text{NoNav}}_i(t)\rangle.
\eea
Fig.~\ref{fig:gapvstH2} shows the energy gap $\Delta E=E_1(t)-E_0(t)$ (\van)
and $\Delta E^{\text{NoNav}}=E^{\text{NoNav}}_1(t)-E^{\text{NoNav}}_0(t)$ (No Navigator)
for the hydrogen molecule with $T=0.1$. 
One can see that the $\hinter$ increases the energy gap during the entire annealing schedule. However, the energy gap difference ${\Delta E^{\text{NoNav}}/ \Delta E}$ 
in Fig.\ref{fig:gapvstH2} is $\mathcal{O}(1)$ and it is not small enough to explain the
difference in the required annealing time;
$T^{\text{\van}}_{\text{CA}}/T^{\text{Standard}}_{\text{CA}}\sim \mathcal{O}(10^{-2})$. Therefore, more dynamical factors must be involved.

\begin{figure}[ht]
\subfigure[]{\includegraphics[scale=0.55]{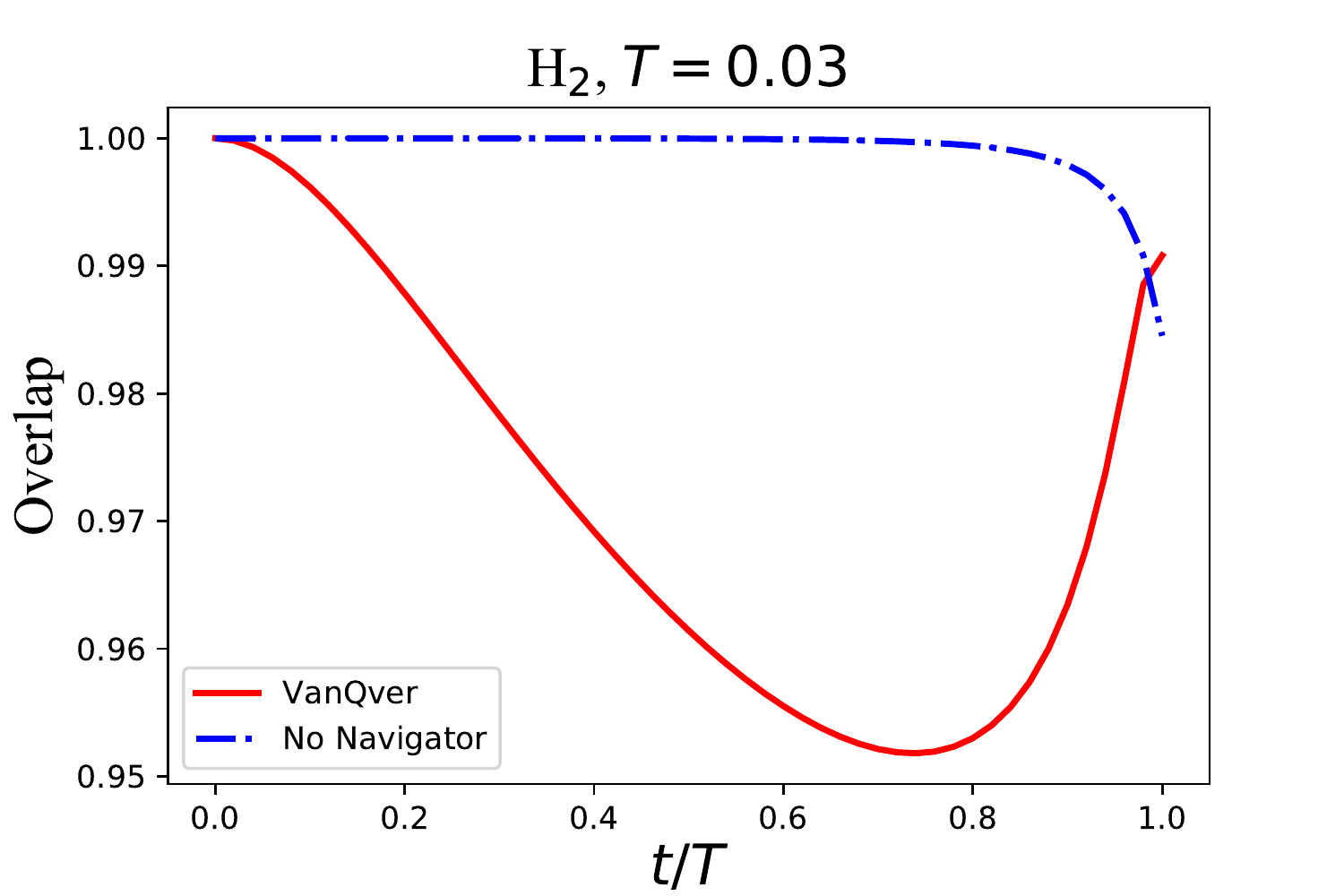} \label{fig:overlapvstH2_T003}}
\subfigure[]{\includegraphics[scale=0.55]{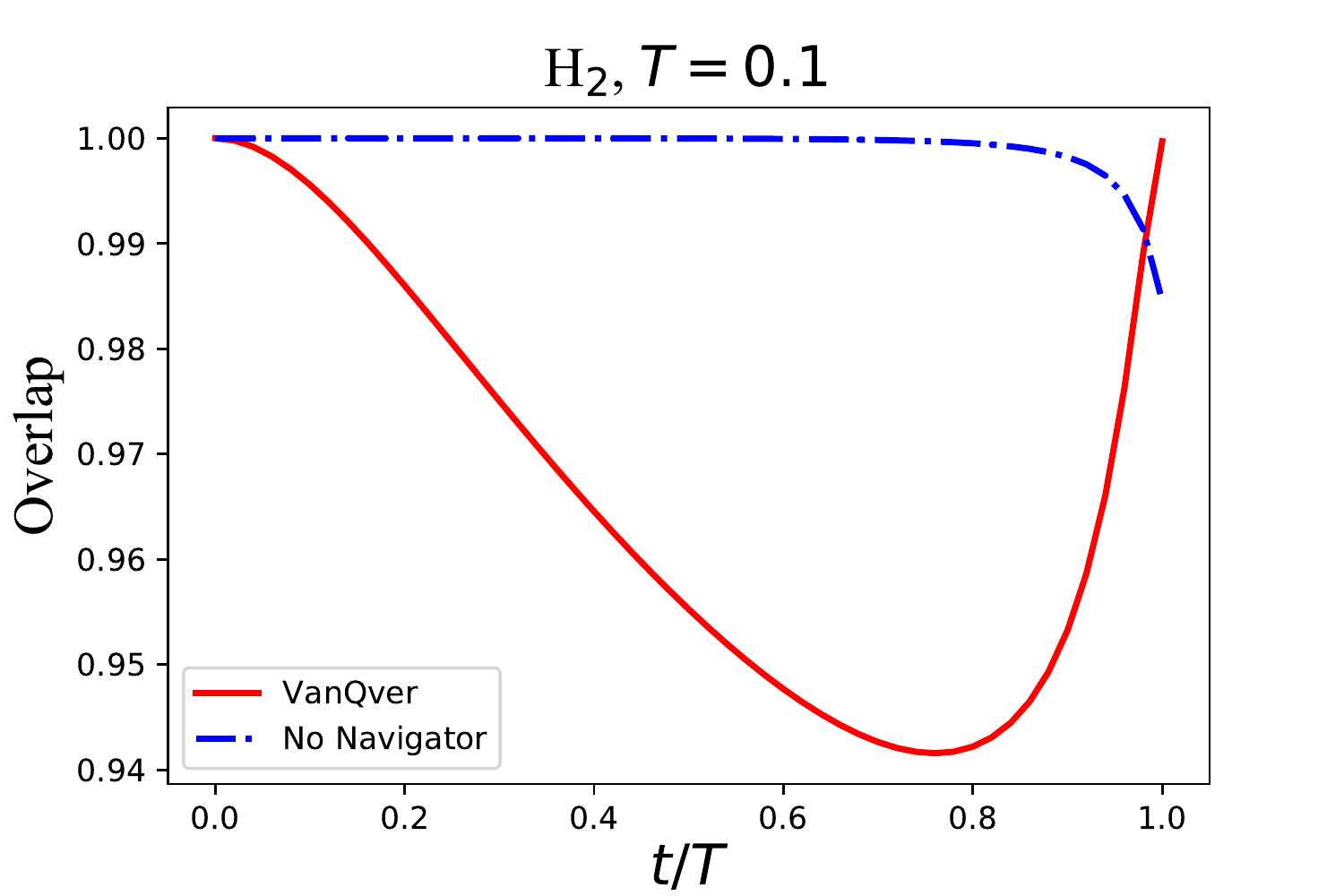} \label{fig:overlapvstH2_T01}}
\caption{Wavefunction overlap between the instantaneous ground state and one generated in the annealing. (a) $T<T_{\text{CA}}^{\text{\van}}$,  (b) $T>T_{\text{CA}}^{\text{\van}}$.
}
\label{fig:overlapvstH2}
\end{figure}

Next, we study the wavefunction overlap.
The wavefunctions generated in the annealing are
\bea
|\psi (t)\rangle 
=\mathcal{T}\exp\left(-i\int_{0}^{t}
H_{(\bf{\eta}^{\text{final}},\bf{\theta}^{\text{final}})}(s)ds
\right)|\psi^{(0)}\rangle \ ,
\eea
for \van~and 
\bea
&&|\psi^{\text{NoNav}} (t)\rangle 
=\mathcal{T}\exp\left(-i\int_{0}^{t}
H_{(\bf{\eta}^{\text{final}},\bf{\theta}=\bf{0})}(s)ds
\right)|\psi^{(0)}\rangle \ , \cr
&&
\eea
without $\hinter$.
In order to understand how close these states are to the instantaneous ground states,
we compute the overlaps,
$|\langle \phi_0(t) |\psi (t)\rangle |$
and
$|\langle \phi^{\text{NoNav}}_0 (t)|\psi^{\text{NoNav}} (t)\rangle|$.
The results for the hydrogen molecule with $T=0.03$
and $T=0.1$ are shown in Fig.~\ref{fig:overlapvstH2}.
The time to chemical accuracy $T_{\text{CA}}^{\text{\van}}=0.048$ sits between these two values.
The wavefunction overlap in \van~becomes very close to 1 for $T>T_{\text{CA}}^{\text{\van}}$ (Fig.~\ref{fig:overlapvstH2_T01}).
Interestingly, the overlap in \van~decreases in the middle of the annealing while it increases towards the end of the annealing.
On the other hand, in the case of no 
$\hinter$, it decreases monotonically and in particular it drops rapidly near the end of the annealing. 
The overlap in \van~takes smaller values except near the end of the annealing.
This means that \van~partially uses excited states in the middle of the annealing so that the wavefunction fully comes back to the ground state at the end of the annealing
where the energy gap becomes small. 
It is also interesting to notice that the minimum of the overlap with the ground state becomes smaller for $T=0.1$ than $T=0.03$ in \van.
Similar features are observed in other molecules.

\section{Conclusion}

A hybrid quantum-classical algorithm \van~was proposed and its efficiency was tested for the purpose of molecular energy estimation. The algorithm is essentially a variational quantum eigensolver that uses adiabatic evolution instead of a gate-based implementation for the state preparation. The adiabatic evolution, however, has been  implemented using parametric Hamiltonians. Namely, a parametric initial driver Hamiltonian, the final Hamiltonian describing the system under study and a parametric navigator Hamiltonian that increases the overlap of the final state with the desired outcome (in this case the state with the lowest eigenvalue). The amplitude of the navigator Hamiltonian is prominence-like, that is, it is gradually increased up to a point during the annealing process and then decreased such that at the end of adiabatic evolution, the only Hamiltonian with a non-zero amplitude is the final Hamiltonian. 
As with the choice of ansatz in gate model VQE, the choice of  parametric navigator Hamiltonian has a critical impact on the performance of this method. In the context of molecular energy estimation, we suggest using a hermitian cluster operator inspired by unitary coupled-cluster theory. 

As a measure of efficiency, the interdependence of annealing time and chemical accuracy were considered. The required annealing times for \van~were found to be significantly shorter than those of standard AQC. Although the shorter annealing time renders this algorithm more amenable to noisy near-term quantum hardware, it is yet to be determined if the shorter single-iteration runtime of variational algorithms also provides an  advantage in terms of overall computational effort when scaling beyond near-term quantum computing technologies. 

Additionally, when a quantum device is coupled to the environment, computational results are not monotonically improved as the annealing time 
increases. Therefore, it is important to analyze the performance of \mbox{\van} and the standard AQC in the presence of noise. We look to address these issues in future work.

It should be noted that \van~can also be used to solve optimization problems. In this case, the final Hamiltonian will be diagonal in the computational basis and will therefore represent a classical energy function. Possible future work expanding on this research would be to determine the optimal choice of the navigator Hamiltonian to achieve shorter annealing time requirements for classical optimization problems.
Ideas already exist in this respect, such as using a non-stoquastic Hamiltonian or an inhomogeneous transverse field.
In some cases, taking non-adiabatic paths is much more efficient than taking adiabatic paths.
Although it is difficult to determine which strategy to employ to shorten the time to solution, 
\mbox{\van}~may be able to systematically survey various strategies.

\section{Acknowledgement}
S.~M. appreciates the hospitality of H.~Nishimori and T.~Takayanagi during his stay at the Tokyo Institute of Technology and the
Yukawa Institute for Theoretical Physics.
 He also appreciates useful discussions with D.~Lidar and P.~Ronagh. 
 We thank Y.~Kawashima, P.~Verma, and S.~Buck  for valuable comments on a draft of the manuscript. 
 We thank Marko Bucyk for reviewing and editing the manuscript, and J.~Loscher and A.~Saidmuradov for technical support.

\appendix
        
\section{Molecules used in the numerical experiments}
\label{App:geometries of molecules}
\begin{figure}[ht]
\includegraphics[scale=0.3]{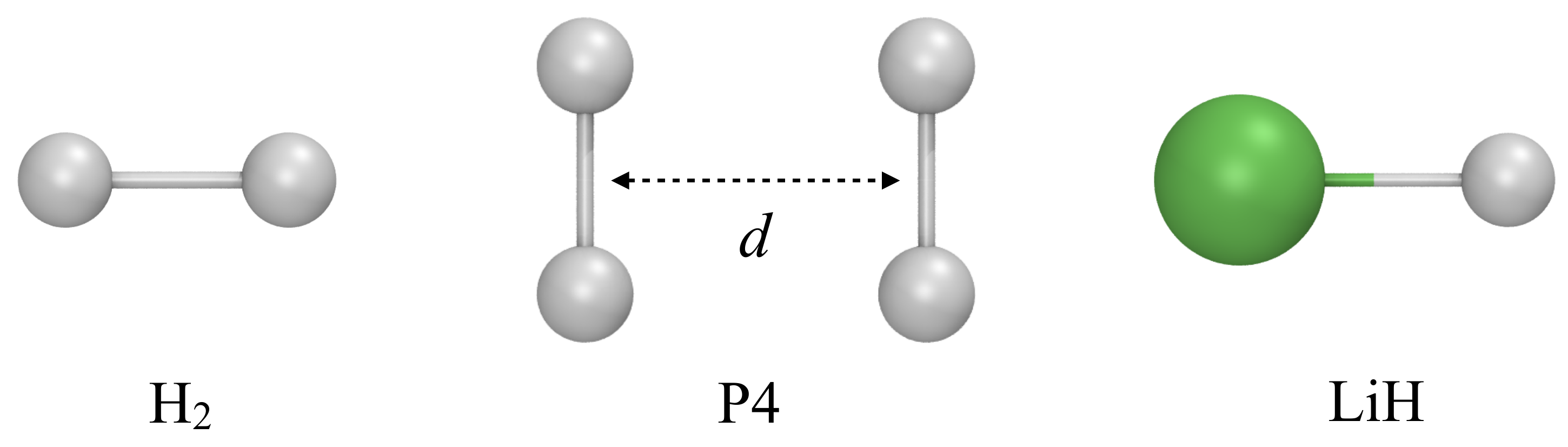}  
\caption{The geometry of
H$_2$, P4, and LiH. 
 }
 \label{fig:mols}
\end{figure}

In the main text, we considered H$_2$, P4, and LiH. The geometries of these molecules are shown in Fig.~\ref{fig:mols}.
The nuclear separation distance for H$_2$ and LiH is $1$ $\ang$. 
For P4, the intermolecular separation distance for each hydrogen molecule is $2$ $\ang$, and $d$ represents the separation distance of the two hydrogen molecules. We investigated $d$ between $0.4$ $\ang$ and $4$ $\ang$.

\section{\van~algorithm}
\label{App:algorithm}
 \setcounter{algocf}{0}
\begin{algo} 
\caption{\,\,\textit{VanQver}}
\textbf{Step 1}. Determine the time profiles $A(t), B(t),$ and $C(t)$ and an initial and navigator Hamiltonians $\hini (\bf{\eta} )$ and  {$\hinter(\bf{\theta} )$} to define the
 dependent Hamiltonian $H_{(\bf{\eta} ,\bf{\theta} )}(t)=A(t)\hini (\bf{\eta} )+B(t)\hfin+C(t)\hinter(\bf{\theta} )$.

\textbf{Step 2}. Set an annealing time $T$ and the initial  and  navigator variational parameters $\bf{\eta} =\bf{\eta} _0$ and $\bf{\theta} =\bf{\theta} _0$ for a quantum device.
Set the tolerance for termination $\epsilon_{tol}$ for a classical optimizer.

\textbf{Step 3}. Run the AQC algorithm on a quantum device with the time-dependent Hamiltonian $H_{(\bf{\eta} ,\bf{\theta} )}(t)$
to generate a quantum state state 
$|\psi_{(\bf{\eta} ,\bf{\theta} )}(T)\rangle=\mathcal{T}exp(-i\int_{0}^{T} H_{(\bf{\eta} ,\bf{\theta} )}(t) dt)|\psi(0)\rangle $.

\textbf{Step 4}. Estimate the expectation value of the final Hamiltonian $E=\langle\psi_{(\bf{\eta} ,\bf{\theta} )}(T) |\hfin|\psi_{(\bf{\eta} ,\bf{\theta} )}(T)\rangle$. 

\textbf{Step 5}. Send $(E,\bf{\eta} ,\bf{\theta} )$ to a classical computer and use a classical optimizer to generate a new set of values for variational parameters $(\tilde{\bf{\eta}} ,\tilde{\bf{\theta}} )$. 

\textbf{Step 6}. Send $(\tilde{\bf{\eta}} ,\tilde{\bf{\theta}} )$ to the quantum hardware.

\textbf{Step 7}. Repeat \textbf{Step 3} through \textbf{Step 6} until the energy $E$ converges with $\epsilon_{tol}$.

\textbf{Step 8}. Output the energy $E$.
\end{algo}

Detailed steps of \van~are shown in Algorithm 1. In the case of our  numerical simulations, in \textbf{Step 1} we chose $A(t)=1-\left({t\over T}\right)^2, B(t)=\left({t\over T}\right)^2$,  and $C(t)= {t\over T}\left(1-{t\over T}\right)$.

In \textbf{Step 4} of Algorithm 1, there is a slight difference between combinatorial optimization problems and quantum simulations.
In the case of combinatorial optimization problems, $\hfin$ is a classical Hamiltonian and the final state is a classical state.
Therefore, measuring all the qubits in the computational basis will give an expectation value of $\hfin$.
In the case of quantum simulations, terms in $\hfin$ do not commute with each other and the final state will be an entangled state.
Therefore, the number of required measurements will increase significantly.
Let  $\hfin=\sum_{k}h_k\sigma^{k}_1\otimes \sigma^{k}_{2}\otimes\cdots \sigma^{k}_{N}$, where $\sigma^{k}_{i}\in [\sigma^{x},\sigma^{y},\sigma^{z},\mathbb{I}]$.
We group the terms into mutually commuting contributions, so the terms of each group can be measured simultaneously.
If the required measurement for grouped terms is not in the computational basis, a change of basis is needed which can be implemented using single-qubit rotations so that all the terms 
are functions of only $\sigma^{z}$ or $\mathbb{I}$. 
For instance, $\mathbb{I}\otimes \mathbb{I}\otimes \sigma^{x}\otimes\sigma^{x}$ and $\mathbb{I}\otimes \sigma^{x}\otimes\sigma^{x}\otimes \mathbb{I}$
commute with each other. Therefore, they are included in the  same group for the purpose of measurement. In order to perform a measurement in the computational basis, we need to rotate the second, third, and fourth qubits from the $x$ direction to the $z$ direction.
This idea was first suggested by~\cite{Hardware-efficientIBM} in the context of the gate model VQE. 

The algorithm, in addition to being a hybrid quantum--classical algorithm, is also a hybrid of adiabatic evolution and gate model quantum computation. Instead of using a parametric circuit of quantum gates, the state preparation step is implemented using a parametric adiabatic evolution. Once the state preparation has been performed, we use an expectation estimation approach borrowed from gate model quantum computing to obtain an estimate of the energy of the current state.

\section{Numerical results for  H$_2$ and LiH}
\label{app:Numerical results for  H2 and LiH}
\begin{figure*}[ht]
\subfigure[]{\includegraphics[scale=0.43]{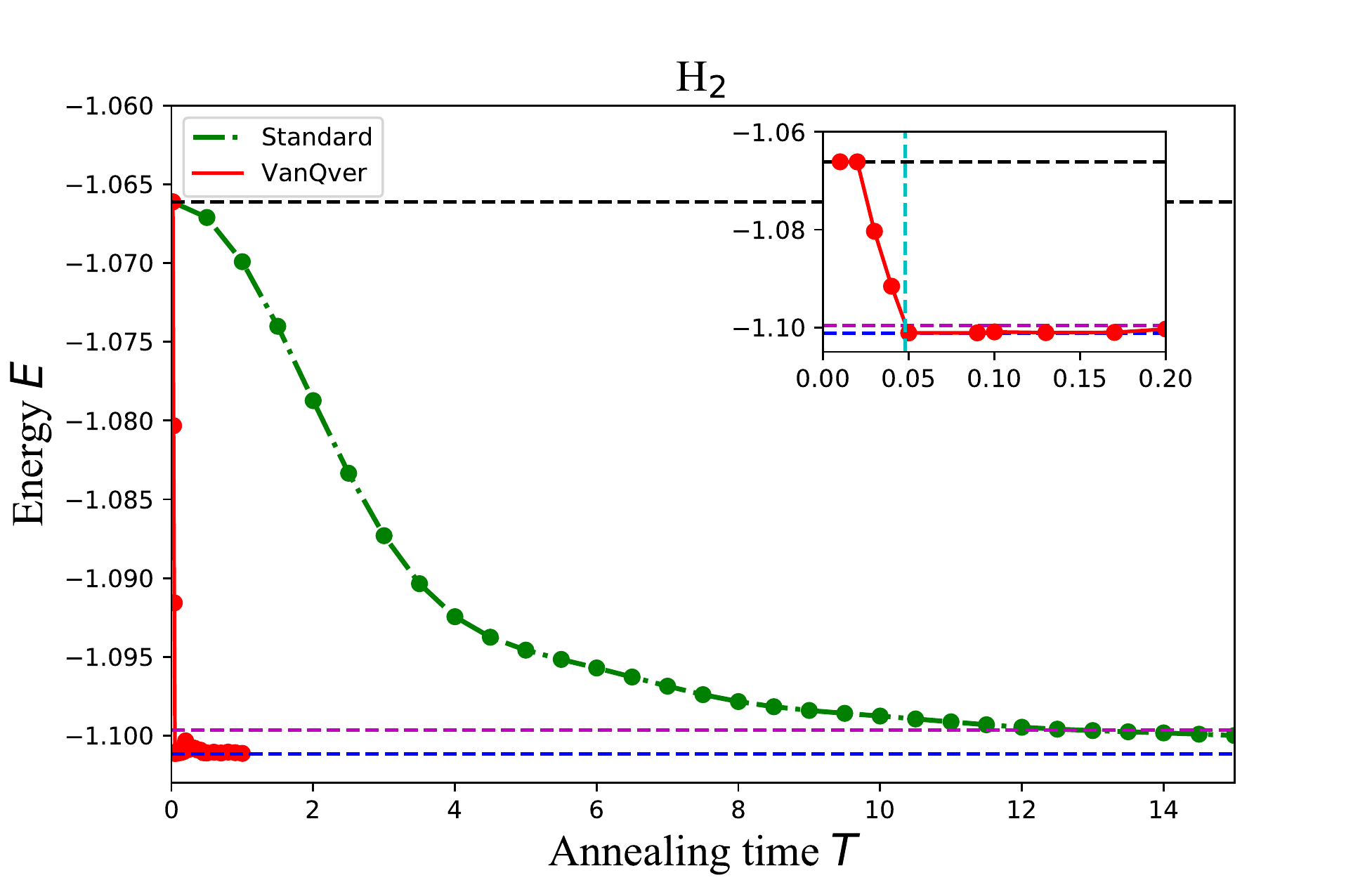}  \label{fig:EvsT_H2}}
\subfigure[]{\includegraphics[scale=0.38]{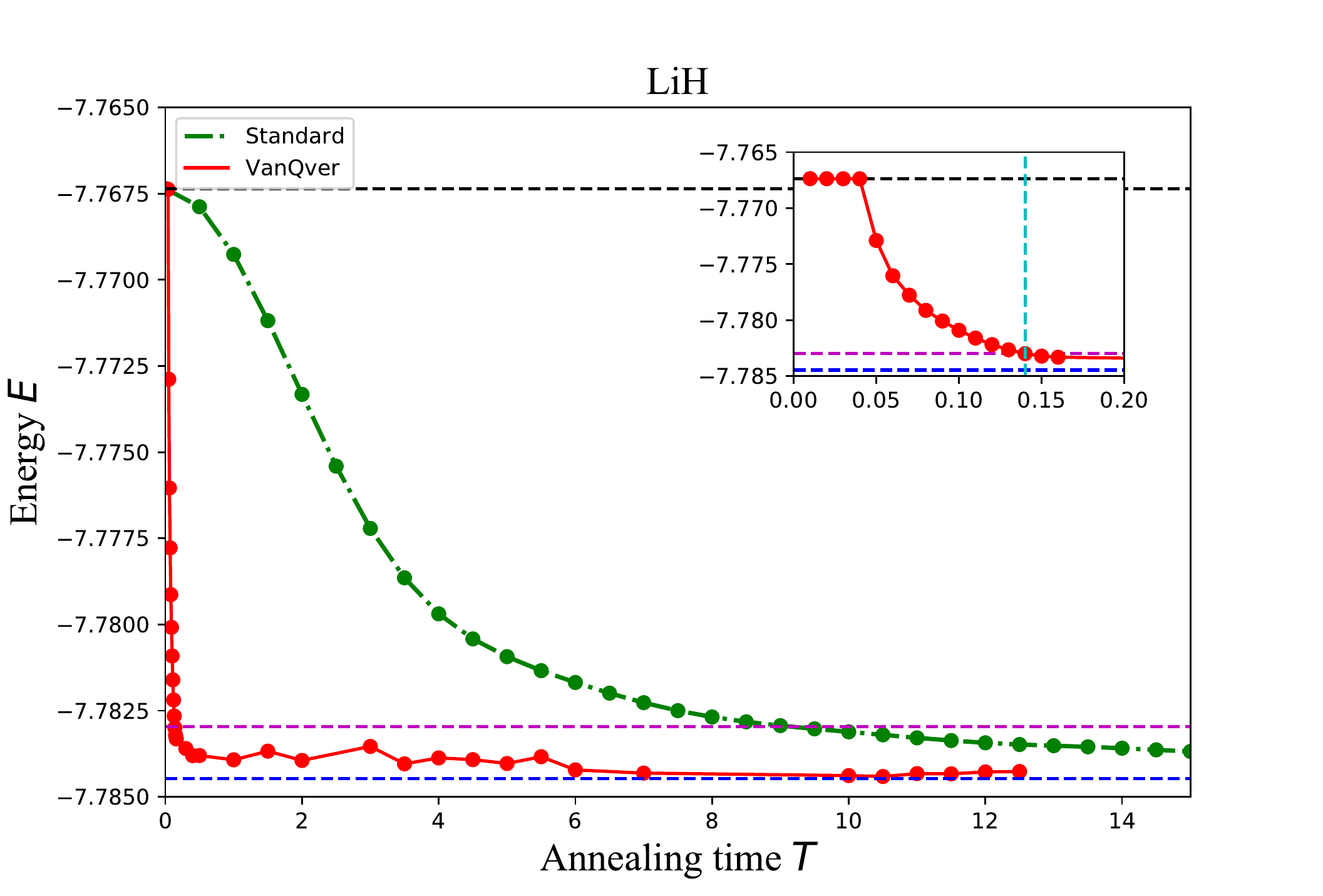} \label{fig:EvsT_LiH} }
\caption{The energy and annealing time for H$_2$ and LiH.
The red points represent the results obtained using \van, the green points represent the results obtained using standard AQC with $\hini=H_{\text{MP}}$, and the dotted lines in blue and magenta represent
the exact and chemical accuracy energies, respectively.
The dotted line in light blue in the insets represents $T^{\text{\van}}_{\text{CA}}$.
The tolerance is $\epsilon_{\text{to;}}=0.001$.
}
\label{fig:EvsT-SM}
\end{figure*}

In the main text, we show the results for P4. The results for H$_2$ and LiH are shown in Fig.~\ref{fig:EvsT-SM},
in which the energy $E=\langle\hfin\rangle$ is a function of the annealing time $T$ for H$_2$ (Fig.~\ref{fig:EvsT_H2})
and LiH (Fig.~\ref{fig:EvsT_LiH}).
In standard AQC, the annealing time to chemical accuracy is $T^{\text{Standard}}_{CA}=13$ for H$_2$ and $T^{\text{Standard}}_{CA}=9.5$ for LiH.
This shows that the difficulty of the computation depends not only on the number of qubits used to represent the molecule but also the distance between nuclei.
In \van, $T^{\text{\van}}_{CA}=0.048$ for H$_2$ and $T^{\text{\van}}_{CA}=0.14$ for LiH.


\section{Classical optimizer dependence}
\label{Classical optimizer dependence}

Although the main claim of this work is that VanQver allows us to reach chemical accuracy in energy estimation in a shorter annealing time than standard AQC, it nevertheless relies on classical optimizers to converge to the solution. These optimizers determined what level of accuracy was achievable in our experiments and how many iterations were necessary to reach it. In our code, we used the library QuTip to simulate AQC and the BFGS solver from the SciPy library function scipy.minimize as a classical optimizer.

\begin{figure}[ht]
\subfigure[]{\includegraphics[scale=0.4]{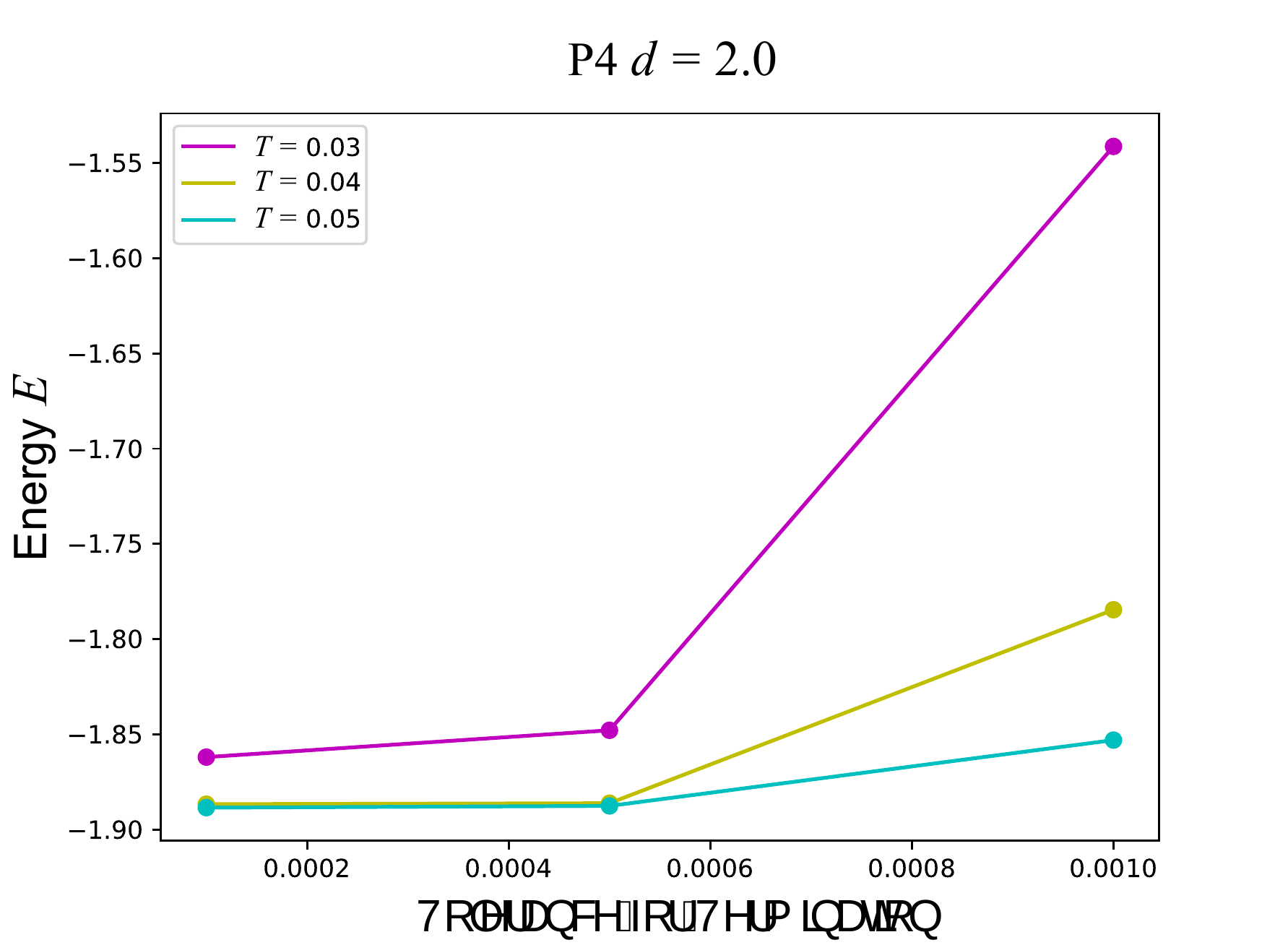} \label{fig:EvsTol_P4-20}}
\subfigure[]{\includegraphics[scale=0.44]{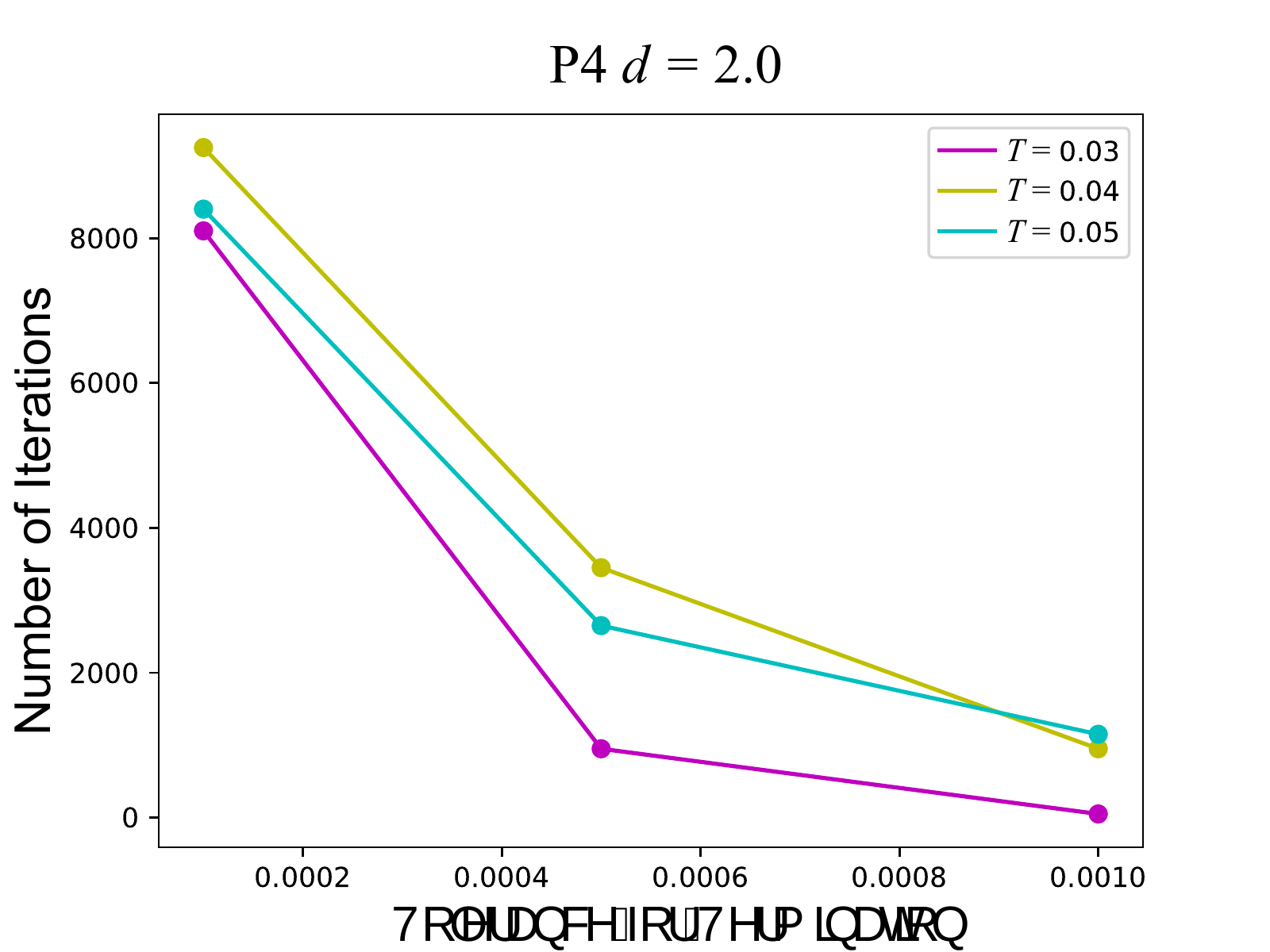} \label{fig:ItevsTol_P4-20}}
\label{Tolerance-E}
\caption{The tolerance for the termination dependence of the final expectation value of $E$ for $T=0.03, 0.04$, and $0.05$, for P4 with $d=$ 2.0.}
\end{figure}

For a given annealing time $T$, the obtained energy $E$ depends on the tolerance for termination $\epsilon_{tol}$ set for the classical optimizer. We compare the quality of the solution obtained after short annealing times $T=0.03, 0.04$, and $0.05$ with different tolerances $\epsilon_{\text{tol}}=0.001,0.005$, and $0.0001$ in Fig.~\ref{fig:EvsTol_P4-20}.
The figure shows that the tolerance constraint $\epsilon_{tol}$ had a significant impact on the final energy found. However, setting this parameter to a smaller value incurred an increased computational cost. Fig.~\ref{fig:ItevsTol_P4-20} shows that the computational cost induced by decreasing $\epsilon_{tol}$ became more important with smaller values of $\epsilon_{tol}$, as reflected by the number of iterations required. For a sufficiently small $\epsilon_{tol}$, further increases in accuracy would impose a severe computational overhead.

\begin{figure}[ht]
\subfigure[]{\includegraphics[scale=0.5]{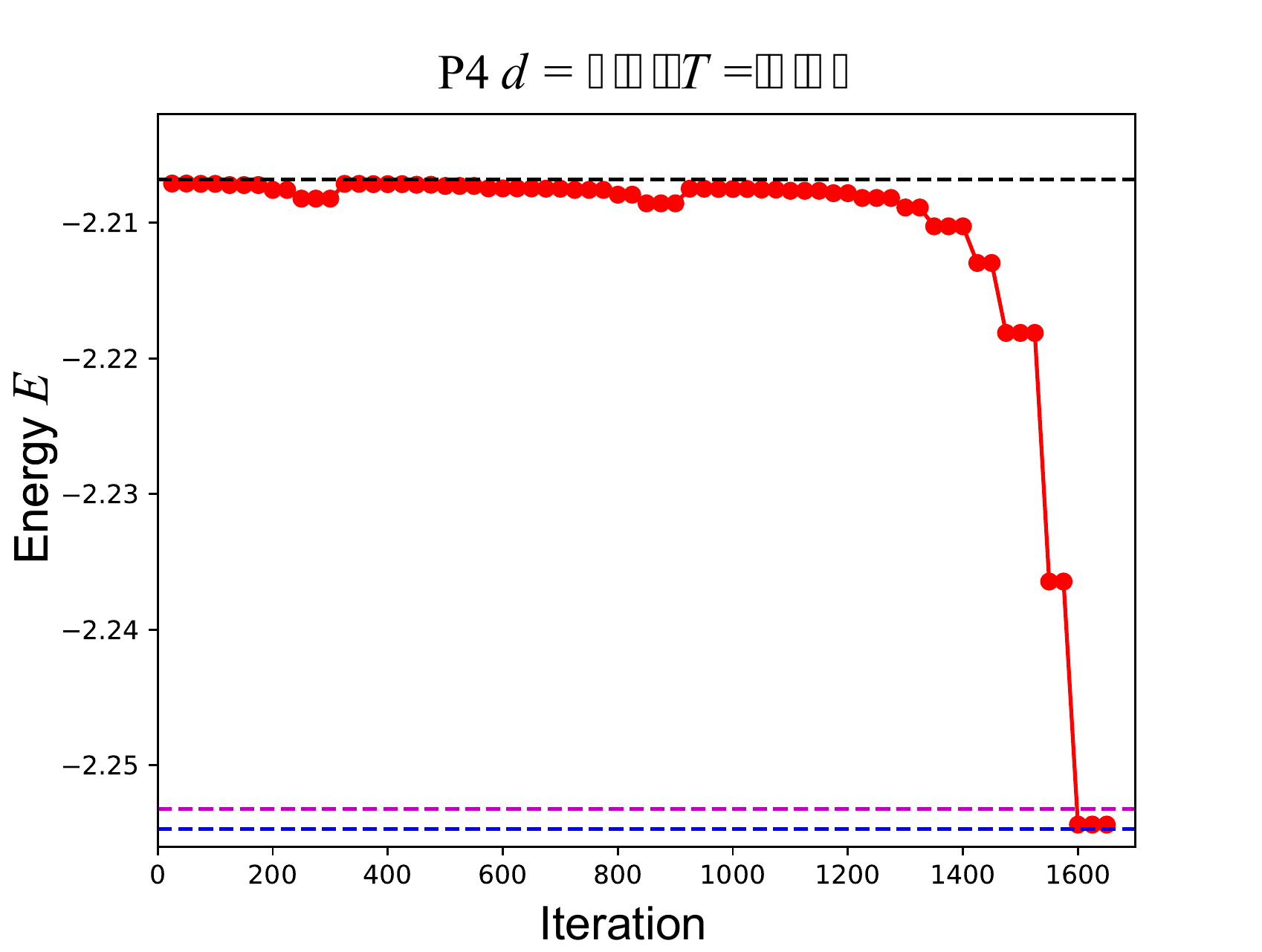}}
\caption{The energy $E$ vs. the number of optimizer iterations for P4 at $d=$0.8 and $T=0.09$, with  $\epsilon_{\text{tol}}$=0.001.
It is noteworthy that our implementation, based on a  restarting strategy used to circumvent a memory issue in QuTip, may have increased the number of iterations required to reach chemical accuracy. The kinks around 300 and 700 iterations represent where the the restarting occurred. 
The graph spans the entire computational effort.}
\label{fig:EvsIterationP408T009}
\end{figure}

Fig.~\ref{fig:EvsIterationP408T009} plots the energy of the system against the number of iterations, in this case for P4 with $d=0.8$ and an annealing time of $T=0.09$. The classical optimizer spent the first 1000 iterations exploring parameters, returning an energy value close to the HF energy,  before dropping quickly to a lower-energy state within chemical accuracy after about 1650 iterations, showing that it may take a long time for the optimizer to find the correct direction to update variational parameters in the parameter space.
Convergence can be improved by conducting a sampling of the energy surface in parallel in order to quickly identify a promising direction for the search, before further iteration.

\begin{figure}[ht]
\subfigure[]{\includegraphics[scale=0.5]{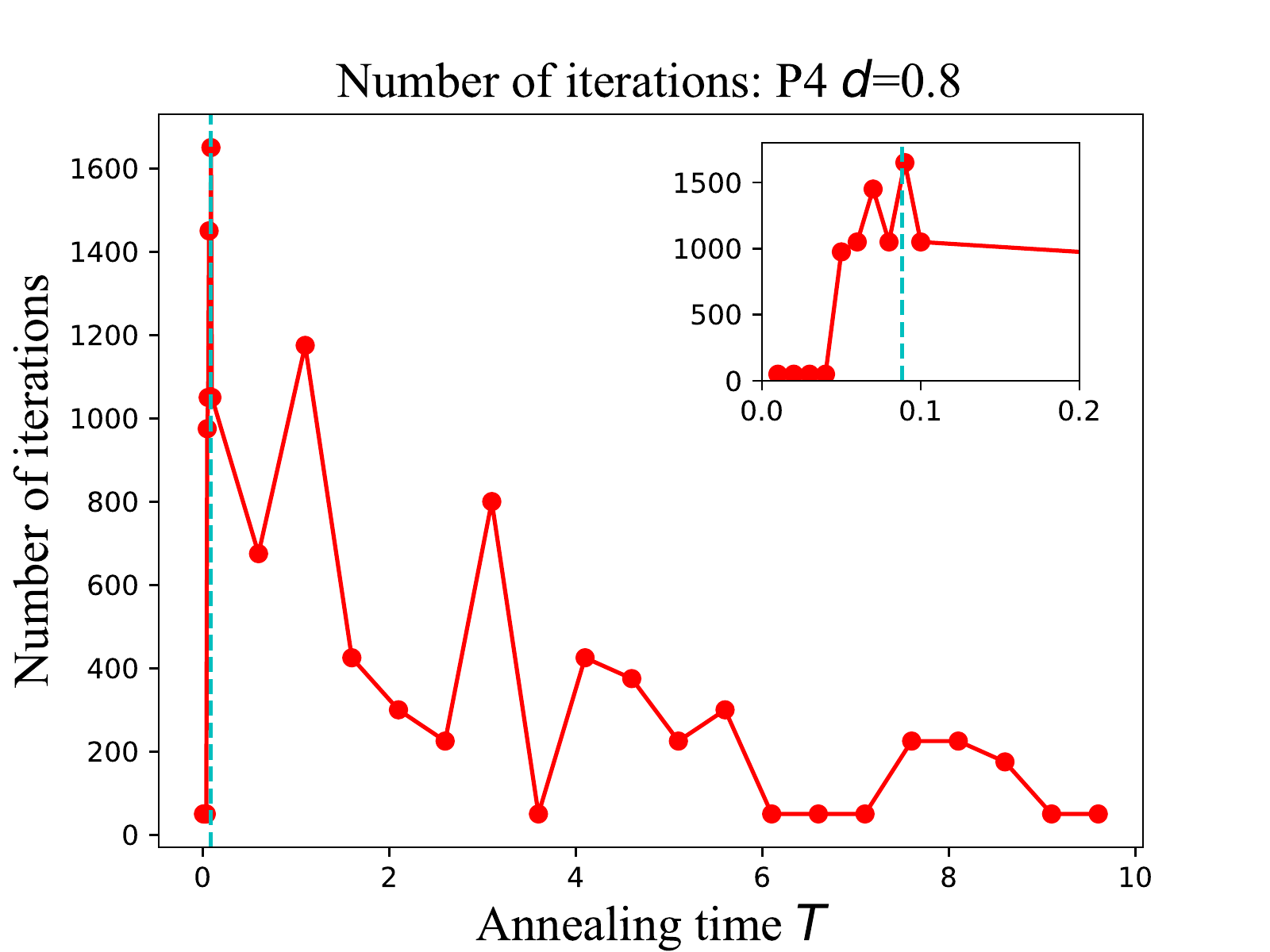}}
\caption{The number of iterations before convergence vs. annealing time for P4 at $d=0.8$. The tolerance for termination was set to $\epsilon_{\text{tol}}$=0.001.
The time to chemical accuracy was $T^{\text{\van}}_{CA}=0.088239751$, shown using a dotted line in cyan.
}
\label{fig:IterationvsT}
\end{figure}

Fig.~\ref{fig:IterationvsT} shows the number of iterations as a function of annealing time $T$ for P4 with $d=0.8$.
When the annealing time was too short, the optimizer converged quickly without improving the result, and the obtained energy was the HF energy.
Increasing the annealing time, we observed dramatic improvements in the quality of the solution over a small window of $T$, between $T$ = 0.05 and 0.09, as shown in Fig.~\ref{fig:EvsT_P4-08}.

This coincides with the peak number of iterations in Fig.~\ref{fig:IterationvsT}, showing that this improvement can be attained at the cost of additional iterations; we noticed that chemical accuracy was reached for $T^{\text{\van}}_{CA}=0.088$. As we further increased the annealing time $T\ge T^{\text{\van}}_{CA}$, we noticed that chemical accuracy was always met and that the number of iterations required to converge tended to decrease.

In order to attain a low-energy state of $\hfin$, 
\van~relies on $\hinter$ to change the annealing path.
We can see that tuning the annealing path had a significant impact
 for shortening annealing times; whereas traditional AQC remained close to the HF energy (see Fig.~\ref{fig:EvsT_P4-08}), $\hinter$ allowed us to attain chemical accuracy after a large number of iterations.
Increasing the annealing time resulted in broadening the range of
annealing paths to obtain the accurate energy, allowing the optimizer to attain convergence in fewer iterations.

\section{The Hartree--Fock Hamiltonian}

From a fermionic Hamiltonian perspective, an initial Hamiltonian may consist of one-electron terms 
\bea
\hini^{\text{ferm}}=\sum_{p}h_{pq}a^{\dagger}_{p}a_{q}\,,
\eea
which includes the HF Hamiltonian.
The form of the initial qubit Hamiltonian used in the numerical experiment
appears under JW transformation when the fermionic Hamiltonian is diagonal in a basis of canonical Hartree-Fock orbitals. 
In our numerical simulations, we employed the canonical RHF Hamiltonian used in~\cite{AQCHmp}.
Equation (\ref{Hmp:H2}) is the HF Hamiltonian for H$_2$ with a nuclei separation distance of $1$ $\ang$:
\bea
H_{\text{MP}}=\sum_{i=1}^{4}h_i \sigma^{z}_{i}+h_{I}\mathbb{I}\,.
\label{Hmp:H2}
\eea
The coefficients are 
$h_{1},h_{2}=  0.2422208402$, 
$h_{3}, h_{4}=-0.2287509695$, and 
$h_{I}=            0.50223746961$.
The first and second qubits represent occupied spin-orbitals and the third and fourth qubits represent virtual spin-orbitals.
Therefore, the signs of the coefficients were appropriately chosen for our purpose.
Note that the eigenvalue of $H_{\text{MP}}$ does not provide the HF energy.

\bibliography{ref}

\end{document}